\documentclass[11pt,a4paper]{article}
\usepackage{jheppub}

\usepackage{graphicx}
\usepackage{dcolumn}
\usepackage{bm}

\usepackage{booktabs}
\usepackage[english]{babel}
\usepackage{amsmath,amssymb,amsbsy,amstext, amsthm, simplewick}
\usepackage{graphicx}
\usepackage{amsfonts}
\usepackage{amssymb}
\usepackage{upgreek}
\usepackage{cancel}
\usepackage{color}
\usepackage{slashed}
\usepackage[dvipsnames]{xcolor}
\usepackage{tikz-cd}
\usepackage{feynmp-auto}
\usepackage{color}
\usepackage{soul}
\usepackage{dsfont}
\usepackage{verbatim}
\usepackage[utf8]{inputenc} 
\usepackage{soul}
\usepackage{epigraph}
\usepackage{todonotes}
\usepackage{csquotes}

\newcommand{\half}{\frac{1}{2}}

\begin{document}

\title{Non-Invertible Peccei-Quinn Symmetry, Natural 2HDM Alignment, and the Visible Axion}

\author[]{Antonio Delgado}
\author[]{\& Seth Koren}
\affiliation[]{Department of Physics and Astronomy, University of Notre Dame, South Bend, IN, 46556 USA}

\date{\today}

\abstract{
We identify $m_{12}^2$ as a spurion of non-invertible Peccei-Quinn symmetry in the type II 2HDM with gauged quark flavor. 
Thus a UV theory which introduces quark color-flavor monopoles can naturally realize alignment without decoupling and can furthermore revive the Weinberg-Wilczek axion. 
As an example we consider the $SU(9)$ theory of color-flavor unification, which needs no new fermions.
This is the first model-building use of non-invertible symmetry to find a Dirac natural explanation for a small \textit{relevant} parameter.
}

\maketitle

\makeatletter
\def\l@subsection#1#2{}
\def\l@subsubsection#1#2{}
\makeatother


\section{Introduction}

Recently we have learned that subtle symmetries which act on both local and extended operators offer us a refinement of the naturalness strategy which can yield concrete model-building guidance. 
The systematic study of symmetries involving extended operators began \textit{en force} a decade ago with the seminal `Generalized Global Symmetries' \cite{Gaiotto:2014kfa}. This expanding paradigm of symmetries has blossomed in recent years with many exotic structures uncovered, but until recently this new insight into the structure of quantum field theories had not been capitalized upon by particle physicists.
This novel framework rephrases symmetries in terms of the existence of defect operators which have topologically invariant correlation functions, and generalize familiar notions of Gaussian surfaces or Noether charges. This change in perspective on symmetries allows us to more-easily generalize, and in particular to find there exist symmetry structures which go beyond symmetry \textit{groups}. 

One way to go beyond group-like symmetries is to consider symmetry structures implemented by non-unitary symmetry defect operators, as has recently been under intense investigation (for reviews see e.g. \cite{Brennan:2023mmt,Schafer-Nameki:2023jdn,Shao:2023gho}) and will be our main tool in this work. The construction of four-dimensional `non-invertible chiral symmetries' was understood only recently \cite{Cordova:2022ieu,Choi:2022jqy}. This has turned out to be an extremely profitable direction for model building, as the intertwining of zero-form and higher-form symmetries into a non-invertible symmetry structure means that on the one hand this symmetry constrains the form of the Lagrangian through its action on local degrees of freedom, and on the other hand it necessarily breaks at higher energies when new degrees of freedom appear in the theory and some IR extended operators are realized as dynamical objects in the UV.\footnote{Another way to go beyond groups is to look at higher-groups \cite{Cordova:2018cvg,Benini:2018reh}, which are a different way symmetries acting on objects of different dimensions can be intertwined. These have been found richly appearing in theories of particle physics, in particular in mixing the zero-form flavor symmetries of the SM with the magnetic one-form symmetry of hypercharge \cite{Cordova:2022qtz}.} 
We refer the reader to Section 7 of \cite{Koren:2024xof} for a pedagogical introduction to the basics of generalized symmetries and their breaking.

These non-invertible symmetry structures offer us a concrete model-building strategy to learn from the IR which operators may be generated by non-perturbative gauge theory effects in a UV completion which introduces the appropriate magnetic monopoles. As an effective field theorist, this is a surprising sort of information to be able to learn from an infrared symmetry analysis, and it offers us further insight into our theories past standard Wilsonian techniques. Indeed, such an analysis in the one Higgs doublet model has led to theories of lepton flavor unification which produce naturally exponentially suppressed Dirac (or Majorana) neutrino masses \cite{Cordova:2022fhg} and to theories of quark color-flavor unification which revive the massless quark solution to strong CP \cite{Cordova:2024ypu}. Here we turn only slightly further beyond the SM to the two Higgs doublet model (2HDM) and ask again what interesting generalized symmetry structures may be present. For the first time we will use non-invertible symmetry to shed light on the small size of a \textit{relevant} parameter, as contextualized in Table \ref{tab:modelbuilding}.

\begin{table}[t]
    \centering
    \renewcommand{\arraystretch}{1.3}
    \begin{tabular}{|c|c|c|}
        \hline & Technically Natural & Unnatural\\\hline
        Marginal & $y_\nu$ (CHKO '22)  & $\bar \theta$ (CHK '24) \\\hline
        Relevant & $m_{12}^2$ (DK here)  & \textit{The} Hierarchy Problem\\\hline
    \end{tabular}
    \caption{The progression of model building utilizing the strategy of non-invertible naturalness. Following the understanding of four-dimensional non-invertible symmetries \cite{Cordova:2022ieu,Choi:2022jqy}, first the technically natural Dirac neutrino yukawas were given a Dirac natural origin in lepton flavor unification \cite{Cordova:2022fhg}. Then the technically unnatural strong CP problem was addressed with the massless quark solution in quark color-flavor unification \cite{Cordova:2024ypu}. Now on to relevant parameters, here we explore a Dirac natural origin for the technically natural $m_{12}^2$ mixing the two Higgs doublets of the 2HDM.}
    \label{tab:modelbuilding}
\end{table}

A second Higgs doublet is a simple, obvious addition to the Standard Model, and such a field is necessarily present in many of our best theories of the universe at very small distances. What is less certain is whether this second Higgs doublet will appear close in mass to our first Higgs doublet, and so play an active role in \textit{nearby} (B)SM phenomenology. At its deepest level the question of how reasonable it is to expect such a degree of freedom to exist likely depends on how \textit{The} Hierarchy Problem is resolved in our vacuum. As it is, we have not yet found out empirically, and we will not here offer further insight into the origins of the electroweak scale per se. But \textit{given} that mass parameter, the question of whether the second Higgs doublet is near the electroweak scale is the question of whether we could be living in the `alignment without decoupling' limit. Data has shown us the discovered Higgs is very Standard Model like, and so we must reside in \textit{some} sort of `alignment' limit where one scalar mass eigenstate couples to the SM species very similarly to the Higgs of the 1HDM. It is this question that we here provide a natural ultraviolet resolution to, which is an explanation of the size of the \textit{other} scalar mass parameter in the 2HDM.

Axions are in no need of a motivating introduction, having grown to dominate particle physics in recent years. Theories of axions have been found to have particularly rich generalized symmetries, including higher-groups \cite{Brennan:2020ehu,Hidaka:2020izy,Hidaka:2021mml,Hidaka:2021kkf,Nakajima:2022feg,Brennan:2023kpw,Anber:2024gis} and non-invertible higher form symmetries \cite{Yokokura:2022alv,Choi:2022fgx,Hidaka:2024kfx,DelZotto:2024ngj}. Understanding generalized symmetry structures in axion theories of particle physics has offered insight into ratios of scales \cite{Brennan:2020ehu}, into charge teleportation processes \cite{Choi:2022fgx}, into their ability to probe Yang-Mills global structure \cite{Cordova:2023ent,Reece:2023iqn,Choi:2023pdp,Cordova:2023her}, and into the quality problem in extra-dimensional setups \cite{Reece:2024wrn,Craig:2024dnl}, among other uses e.g. \cite{Brennan:2023kpw,Aloni:2024jpb,Chen:2024tsx}. 
More generally, over the past few years ideas and technology from generalized symmetries are gradually being utilized in (or towards) a variety of particle physics applications, see e.g.~\cite{Cordova:2018cvg,Benini:2018reh,Cordova:2022qtz,Cordova:2022fhg,Cordova:2024ypu,Cordova:2022rer,Koren:2024xof,
Brennan:2020ehu,Cordova:2022ieu,Brennan:2022tyl,Cordova:2023ent,Cordova:2023her,Brennan:2023kpw,Brennan:2023tae, Brennan:2024iau,
vanBeest:2023dbu,vanBeest:2023mbs,
Davighi:2019rcd,Davighi:2020kok,Davighi:2024zip,Davighi:2024zjp,
Anber:2021upc,Anber:2024gis,
Cheung:2024ypq,
Kobayashi:2024yqq,Kobayashi:2024cvp,Funakoshi:2024uvy,
Cherman:2022eml,Cherman:2023xok,Chen:2024tsx,
Hinterbichler:2022agn,Hinterbichler:2024cxn,
Hidaka:2020izy,Hidaka:2021mml,Hidaka:2021kkf,Yokokura:2022alv,Nakajima:2022feg,Hidaka:2024kfx,
Craig:2024dnl,
Choi:2022jqy,Choi:2022rfe,Choi:2022fgx,Choi:2023pdp,
Arbalestrier:2024oqg,
DelZotto:2024ngj,
Heidenreich:2020pkc,Heidenreich:2021xpr,Heidenreich:2023pbi, Fan:2021ntg,McNamara:2022lrw,Asadi:2022vys,Reece:2023iqn,Aloni:2024jpb,Reece:2024wrn,Garcia-Valdecasas:2024cqn,
Wan:2019gqr,Wang:2020xyo,Wang:2021ayd,Wang:2021vki,Wang:2021hob,Wang:2022eag,Putrov:2023jqi,Wang:2023tbj}.

We here give the first concrete model-building use of generalized symmetries in an axion theory to resolve a naturalness issue within the context of two Higgs doublet models. This allows us to revive the most parsimonious axion model: the \textit{visible} Peccei-Quinn Weinberg-Wilczek axion \cite{Peccei:1977hh,Peccei:1977ur,Weinberg:1977ma,Wilczek:1977pj}, where the axion lives in the scalar fields which provide masses for the SM fermions.  

\begin{figure}
    \centering
    \includegraphics[trim={0 0 0 0},clip,width=0.5\linewidth]{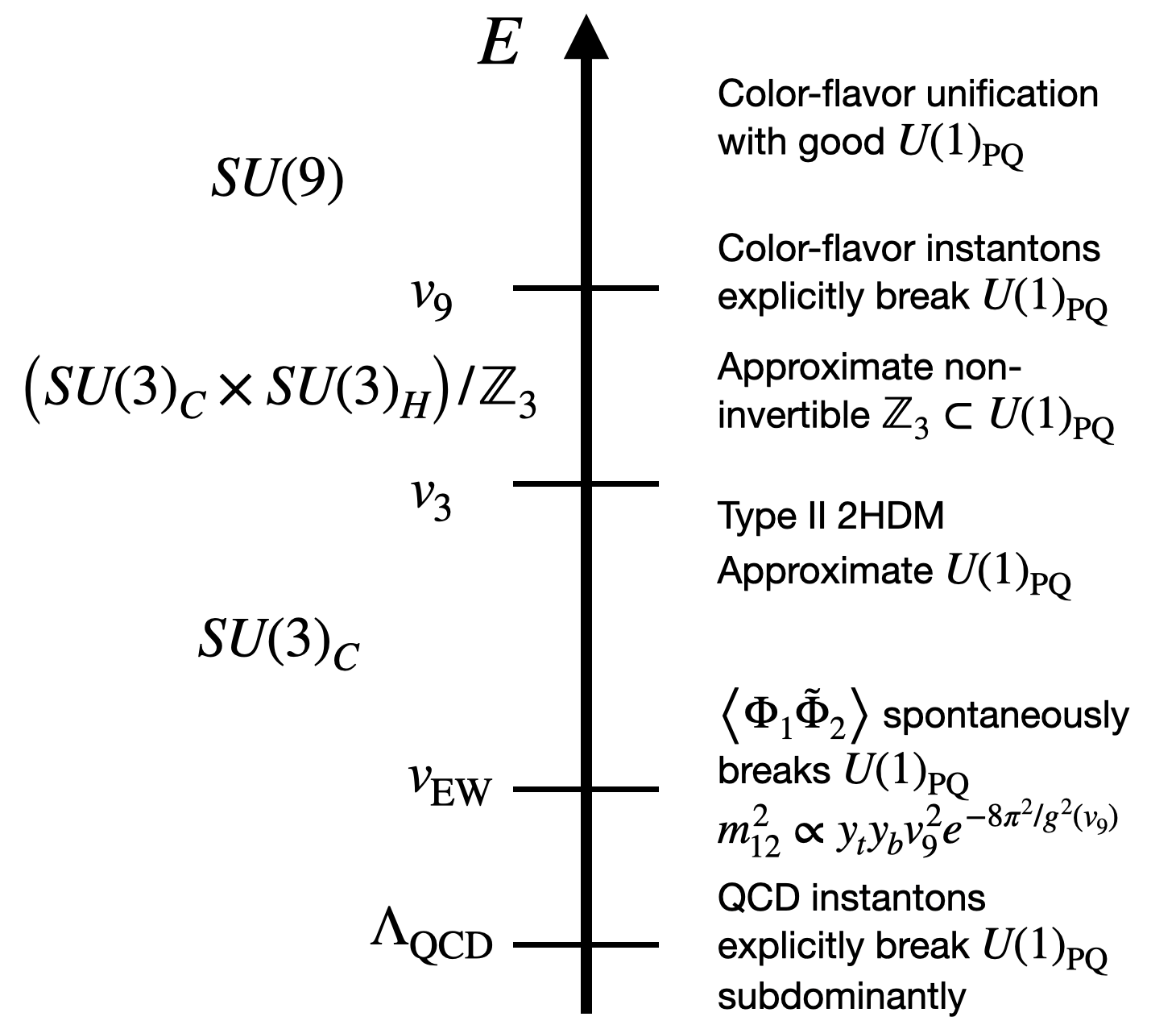}
    \caption{Relevant symmetries and symmetry-breaking. We will proceed in bottom-up fashion beginning in Section \ref{sec:tII2HDM} with the Type II 2HDM phase and then using the non-invertible symmetry in the $SU(3)^2/\mathbb{Z}_3$ phase to discover the $SU(9)$ theory of Section \ref{sec:su9}. In this theory the Peccei-Quinn symmetry is dominantly broken by small color-flavor instantons. }
    \label{fig:energyScales}
\end{figure}

We summarize the symmetries and symmetry-breaking in our model in Figure \ref{fig:energyScales}. In Section \ref{sec:tII2HDM} we will review the Type II 2HDM and its alignment limit. We will see that in a UV embedding of this theory with gauged quark flavor $\frac{SU(3)_C \times SU(3)_H}{\mathbb{Z}_3}$, an important subgroup of the Peccei-Quinn symmetry becomes non-invertible. This guides us to further UV theories in which this subgroup is broken by UV instantons, and we study the $SU(9)$ color-flavor unified theory in particular in Section \ref{sec:su9}. In Section \ref{sec:paramspace} we return to the Type II 2HDM, now as the infrared limit of the $SU(9)$ theory. In Section \ref{sec:A=a} we will discuss this as a model where the 2HDM pseudoscalar is the PQWW axion, and see that there is no quality problem for this visible axion model.

\section{Type II 2HDM and Gauged Quark Flavor} \label{sec:tII2HDM}

The general 2HDM with $\Phi_1, \Phi_2$ having the same electroweak charge assignment has a scalar potential
\begin{align}
    V(\Phi_1, \Phi_2) &= m_{11}^2 \Phi_1^\dagger \Phi_1 + m_{22}^2 \Phi_2^\dagger \Phi_2 - \left[m_{12}^2 \Phi_1^\dagger \Phi_2 + \text{ h.c.}\right] \\
    & + \half \lambda_1 \left(\Phi_1^\dagger \Phi_1\right)^2 + \half \lambda_2 \left(\Phi_2^\dagger \Phi_2\right)^2 + \lambda_3 \left(\Phi_1^\dagger \Phi_1\right) \left(\Phi_2^\dagger \Phi_2\right) + \lambda_4 \left(\Phi_1^\dagger \Phi_2\right) \left(\Phi_2^\dagger \Phi_1\right) \nonumber \\
    & + \left\lbrace \half \lambda_5 \left(\Phi_1^\dagger \Phi_2\right)^2 + \left[\lambda_6 \left(\Phi_1^\dagger \Phi_1\right) + \lambda_7 \left(\Phi_2^\dagger \Phi_2\right)\right]\left(\Phi_1^\dagger \Phi_2\right) + \text{ h.c.}\right\rbrace. \nonumber
\end{align}
The Weinberg-Wilczek theory is a type II 2HDM in which $\Phi_1$ couples to the up-type quarks and $\Phi_2$ to the down-type quarks. 
\begin{equation} \label{eqn:yukawas}
    \mathcal{L} \supset (y_t)^i_j \Phi_1 Q_i \bar u^j + (y_b)^i_j \tilde{\Phi}_2 Q_i \bar d^j.
\end{equation}

Let us consider the classical $U(1)$ symmetries of this theory for generic yukawas. There are 5 free charges for $Q, \bar u, \bar d, \Phi_1, \Phi_2$ and there are two constraints imposed by the yukawa interactions of Equation \ref{eqn:yukawas} which leave $U(1)^3$ remaining symmetry. In this three-dimensional space there is one direction under which the Higgses have the same charge, the gauged $U(1)_Y$, and another direction under which the scalars are uncharged and the fermions are vector-like, $U(1)_B$ (or $U(1)_{B-N_cL}$ in the full theory). Either of these directions has no $SU(3)_C$ anomaly. 

Any direction which is not in the plane spanned by $Y, B$ has a mixed anomaly with QCD and so deserves the moniker of a `Peccei-Quinn symmetry'. 
When such a symmetry is exact, $m_{12}^2, \lambda_5, \lambda_6, \lambda_7 = 0$ all vanish. 
A convenient parametrization is to consider the orthogonal direction to hypercharge for the scalars, which assigns $\Phi_1, \Phi_2$ opposite charges as in Table \ref{tab:inter_charges}. This Peccei-Quinn symmetry has anomaly coefficient $N = - 2 N_g$, where $N_g=3$ is the number of generations of quarks, which means that QCD instantons themselves do not generate any of the PQ-violating relevant or marginal couplings (see Table \ref{tab:spurions}). We will assume that our UV theory has a good such Peccei-Quinn Symmetry, which is manifestly sensible as $M_{\rm pl} \rightarrow \infty$. For finite $M_{\rm pl}$ one should ask about the necessary quality, and we show in Section \ref{sec:quality} that there is no quality problem in this theory as long as the scale of quark flavor physics is not too much higher than current limits.

\begin{table}\centering
\large
\renewcommand{\arraystretch}{1.3}
\begin{tabular}{|c|c|c|c|c|c|}  \hline
 & $SU(3)_C$ & $SU(3)_H$ & $SU(2)_L$ & $U(1)_Y$ & $U(1)_{\rm PQ}$ \\ \hline

${Q}$ & $3$ & $3$ & $2$ & $+1$ & $0$  \\ \hline

${\bar u}$ & $\bar 3$ & $\bar 3$ & $-$ & $-4$ & $1$ \\ \hline

${\bar d}$ & $\bar 3$ & $\bar 3$ & $-$ & $+2$ & $1$ \\ \hline

$\Phi_1$ & $-$ & $-$ & $2$ & $+3$ & $-1$ \\ \hline

$\Phi_2$ & $-$ & $-$ & $2$ & $+3$ & $1$ \\ \hline

\end{tabular}\caption{Symmetry and matter content of the 2HDM with gauged $SU(3)_H$. }\label{tab:inter_charges}
\end{table}

\begin{table}\centering
\large
\renewcommand{\arraystretch}{1.3}
\begin{tabular}{|c|c|}  \hline
 & $U(1)_{\rm PQ}$ \\ \hline
$m_{12}^2$ & $-2$ \\ \hline
$\lambda_5$ & $-4$ \\ \hline
$\lambda_6$ & $-2$ \\ \hline
$\lambda_7$ & $-2$ \\ \hline
\end{tabular}\caption{Spurionic Peccei-Quinn charges for the scalar potential parameters. }\label{tab:spurions}
\end{table}

We will also impose on the tree-level scalar potential a certain $\mathbb{Z}_2$ interchange symmetry, both for simplicity and because it is phenomenologically well-motivated. This $\mathbb{Z}_2$ symmetry will set $\lambda_1 = \lambda_2$ and $m_{11}^2 = m_{22}^2$. Because the PQ symmetry has already set many of the parameters to vanish, there are multiple options for a $\mathbb{Z}_2$ symmetry to impose, most simply $\Phi_1 \leftrightarrow \Phi_2$. 
However the one which commutes with the PQ charge assignment mixes up an interchange with a CP transformation and is often called `GCP2' \cite{Davidson:2005cw,Ferreira:2009wh,Haber:2020wco}:
\begin{equation}
    \text{GCP2}: 
    \begin{pmatrix}
        \Phi_1 \\
        \Phi_2 
    \end{pmatrix} \rightarrow 
    \begin{pmatrix}
        0 & 1 \\
        -1 & 0 
    \end{pmatrix}
    \begin{pmatrix}
        \Phi_1^\star \\
        \Phi_2^\star 
    \end{pmatrix}.
\end{equation}
With the matter content we have, either choice is equivalent. And either choice of symmetry for the tree-level scalar potential is explicitly broken by the yukawas $y_t \gg y_b$ which prevents extending this to $\bar u \leftrightarrow \bar d$ without adding new matter, which we do not pursue here \cite{Draper:2020tyq}.

This \textit{hard} breaking of the symmetry necessarily introduces a perturbation away from the alignment limit, but the shift is loop suppressed as e.g.
\begin{equation}
    \lambda_1 - \lambda_2 \sim  \frac{y_t^4 - y_b^4}{(4\pi)^2} \log\frac{\Lambda}{v_{\rm EW}}
\end{equation} 
and can easily be at the few percent level, which is anyway close enough to the alignment limit to be consistent with all empirical observations thus far. 

Of course this hard breaking of the $\mathbb{Z}_2$ by the yukawas also leads to UV divergent splitting of the masses, which a Wilsonian would estimate as scaling with the cutoff $\Lambda$:
\begin{equation}
    m_{11}^2 - m_{22}^2 \sim \frac{y_t^2 - y_b^2}{(4\pi)^2} \Lambda^2.
\end{equation}
This hard cutoff calculation is an estimate of what a generic UV completion would do, not a dictum. 
Given the matter we have introduced so far, the only finite contribution comes from integrating out the top quark itself and corrects 
\begin{equation}
    m_{11}^2 - m_{22}^2 \sim \frac{m_t^2 - m_b^2}{(4\pi)^2},
\end{equation}
which by contrast is a negligible shift. But this is the same logic as in any discussion of naturalness: We cannot precisely characterize a hierarchy problem until we specify a theory which \textit{predicts} $m_{11}$ and $m_{22}$, and in the absence of such a theory the Wilsonian estimation is useful as a generic expectation.

The philosophy we take in this work is that our theory is not a model to solve \textit{the} hierarchy problem. Then we do not aim to answer the question of the sizes of $m_{11}^2, m_{22}^2$.
Elsewhere we will consider a supersymmetric version of this theory in which it is more reasonable to ask about the sizes of these mass terms as well. 

\subsection{Alignment Without Decoupling and Exceptional Regions}

Two Higgs Doublet Models are a popular way to enlarge the Higgs sector and they have a far richer phenomenology than the SM.
However, the observation of the Higgs boson $h(125)$ and the measurement of its couplings to the SM fermions and vector bosons has drastically impacted the parameter space of 2HDMs. The observed Higgs is extremely SM-like, implying that any 2HDM we realize must reside in an `alignment limit'. The uninteresting way for this to happen is for the extra scalars of the 2HDM to be heavy, which is a decoupling limit where around the electroweak scale the physics is only the SM. This is the case in the MSSM where the alignment limit is achieved for large $m_A$---the reason for this restriction is that supersymmetry relates the values of the quartic couplings of the Higgs doublets to known SM parameters.

On the other hand in a less constrained 2HDM it remains possible that we are in the more interesting case of `alignment without decoupling' \cite{Gunion:2002zf}, where there is a very SM-like Higgs which is not the only scalar around the electroweak scale. The conditions among the different parameters in order to realize alignment without decoupling have been worked out in the literature but one may wonder if those conditions could come from an underlying symmetry instead of being just a particular point in the parameter space subject to uncontrolled renormalization group evolution.

The potential of the 2HDM may accommodate several symmetry structures ($\mathbb{Z}_2$, generalized CP, Peccei-Quinn, $SO(3)$ `Higgs family',...) and various combinations thereof. One can study the relations among the parameters imposed by those symmetries and classify the different families of models, see e.g. Section 5.6 of \cite{Branco:2011iw}. In particular, one can discuss various symmetries which guarantee the alignment limit. 
Much work has been done on low-energy effective field theories with such symmetries. 
Alas, a continuous such symmetry cannot be exact, since it will predict a massless goldstone after EWSB.
Thankfully, there are `exceptional regions of parameter space', identified by \cite{Haber:2021zva}, where this alignment limit may be realized including the effects of soft breaking of the enhanced symmetry. 

However, comparatively little work has been done on ultraviolet theories that realize these structures, and this is what we offer.
The question is: how did this small breaking of the symmetry come about? Studying the low-energy theory with small symmetry-breaking terms as inputs of the theory is ultimately unsatisfying. That is to say, we can do better than technical naturalness---we can try to write theories which are Dirac natural and begin with a good symmetry with small breaking effects generated upon flowing into the infrared. 
Understanding below that the 2HDM Peccei-Quinn symmetry can become non-invertible in quark flavor gauge theories will lead us to UV constructions which predict explicit, calculable, soft breaking by nonperturbative effects.

\subsection{Non-Invertible Peccei-Quinn} \label{sec:NIPQ}

Merely adding the second Higgs doublet to the Standard Model, the Peccei-Quinn symmetry of Table \ref{tab:inter_charges} has a mixed anomaly coefficient
\begin{equation}
    \mathcal{A}_{SU(3)^2_C {\rm PQ}} = - 2 N_g,
\end{equation}
where $N_g=3$ is the number of generations of SM fermions. This leads to an anomalous conservation equation for the $U(1)_{\rm PQ}$ current 
\begin{equation}
    \partial_\mu J^\mu_{\rm PQ} = \frac{\mathcal{A}_{SU(3)^2_C {\rm PQ}}}{32\pi^2} \text{Tr }F_C \tilde{F}_C,
\end{equation}
where $F_C$ is the color field strength, $\tilde{F}_C$ is its Hodge dual. Integrating this we find a change in the PQ charge of
\begin{equation}
    \Delta Q_{\rm PQ} \equiv Q_{\rm PQ}(t = +\infty) - Q_{\rm PQ}(t = - \infty) = \mathcal{A}_{SU(3)^2_C {\rm PQ}} \mathcal{N}_C,
\end{equation}
where $\mathcal{N}_C$ is the instanton number for a given $SU(3)_C$ configuration. 

As is familiar, the $SU(3)_C$ instanton number takes values in the integers, $\mathcal{N}_C\in \mathbb{Z}$.
Then as a result of this anomaly, the $U(1)_{\rm PQ}$ is explicitly broken by QCD instantons, and only a $\mathbb{Z}_{2N_g}$ subgroup remains anomaly-free. Resultingly the QCD instantons do not contribute to the PQ-violating operators in the renormalizable scalar potential, since they can only change the PQ charge by $2 N_g$. This means the first operator they generate is $a_6 (\Phi_1^\dagger \Phi_2)^3$, where one expects $a_6 \sim \Lambda_{\rm QCD}^{-2}$ by dimensional analysis but the effects of these IR instantons are not under analytic control.

Quite generally, a well-motivated possibility is that the approximate global flavor symmetries we see in the SM have arisen from gauged flavor symmetries which are Higgsed, which we take to occur at the scale $v_3\gg v_{\rm EW}$.  One of us discussed in \cite{Cordova:2024ypu} that subtle non-invertible symmetries may appear when quark flavor symmetries are gauged such that there is some non-trivial global structure between $SU(3)_C$ and the flavor gauge group\footnote{For discussions of particle physics implications of gauge theories of groups with different global structures see e.g.  \cite{Tong:2017oea,Koren:2024xof,Cordova:2023her,Choi:2023pdp,Reece:2023iqn,Anber:2021iip,Li:2024nuo,Alonso:2024pmq,Hsin:2024lya,Dierigl:2024cxm}.}
\begin{equation}
    \frac{SU(3)_C \times U(1)_{B_1 + B_2 - 2B_3}}{\mathbb{Z}_3} \quad \text{ or } \quad \frac{SU(3)_C \times SU(3)_H}{\mathbb{Z}_3},
\end{equation}
where the former is flavored baryon number and the latter is non-Abelian gauged quark flavor. The non-invertible Peccei-Quinn symmetry we discuss here is closely analogous to the symmetry in that work, but extended from the 1HDM to the 2HDM. Let us emphasize that this is something special that can be done because of the `coincidence' that the SM has the same number of colors of QCD and generations of fermions, $N_c=N_g$.

We describe in brief the non-invertible symmetry which appears in the $SU(3)^2/\mathbb{Z}_3$ 2HDM theory and refer to \cite{Cordova:2024ypu} for further details on the construction of the non-invertible symmetry and on the other case. In $SU(3)_C$ the instanton number is integer valued, but not so in $SU(3)_C/\mathbb{Z}_3$, where we quotient by the center of the group. Here the instanton number is only necessarily integer valued in a spacetime with trivial topology and no 't Hooft operator insertions. On a \textit{general} manifold, or in the presence of 't Hooft lines, the instanton number in $SU(3)/\mathbb{Z}_3$ can become fractional, 
\begin{equation}
    SU(3)_C: \mathcal{N}_C \in \mathbb{Z}, \qquad SU(3)_C/\mathbb{Z}_3: \mathcal{N}_C \in \mathbb{Z}/3.
\end{equation}
The group structure in our case is more complicated because it is a diagonal center we are quotienting by, but at the level of the instanton numbers we can think about both $\mathcal{N}_C, \mathcal{N}_H \in \mathbb{Z}/3$ with the restriction
\begin{equation}
    \mathcal{N}_C = \mathcal{N}_H \ (\text{mod 1}).
\end{equation}
A useful physical perspective on this can be gained through constructing $SU(3)^2/\mathbb{Z}_3$ by beginning with a $SU(3)^2$ gauge theory and then gauging the diagonal electric one-form symmetry $\mathbb{Z}_3^{(1)}$, see e.g. \cite{Brennan:2023mmt} for insight into this procedure. Gauging this electric one-form symmetry restricts the allowed electric representations in this theory, and in line with a general notion of Dirac quantization this means that the theory should gain magnetic representations. Indeed, the $SU(3)^2/\mathbb{Z}_3$ theory has 't Hooft lines carrying both color-magnetic and flavor-magnetic charges and enjoying a $\mathbb{Z}_3^{(1)}$ magnetic one-form symmetry.

How should we think about these `exotic' fractional instantons as low-energy effective field theorists? 
These instantons can be realized whenever there are 2-cycles around, as for example if spacetime is $S^2 \times S^2$. 
Then if we should analyze the symmetries of the theory by looking at the partition function on a general manifold, one might think the fractional instanton numbers explicitly break more of the PQ symmetry which remained after taking into account the integer instantons. In particular the counting is now 
\begin{equation}
    \Delta Q_{\rm PQ} = - 2 N_g \mathcal{N}_C - 2 N_c \mathcal{N}_H,
\end{equation}
and for a minimal fractional instanton $\mathcal{N}_C = \mathcal{N}_H = 1/3$ we see any PQ rotations that are not in a $\mathbb{Z}_4$ subgroup have become anomalous. Combined with the integer instanton backgrounds, then, we would conclude $U(1)_{\rm PQ} \rightarrow \mathbb{Z}_6 \cap \mathbb{Z}_4 = \mathbb{Z}_2$.

However it is also true that in $\mathbb{R}^4$ without magnetic monopoles, where we live to good approximation, these instanton numbers cannot be realized. If we were to look at this theory defined strictly on empty $\mathbb{R}^4$ then we would find that the S-matrix continues to preserve the $\mathbb{Z}_6$ which is preserved by the integer instantons. This seems to give some ambiguity as to the status of PQ---should we think the fractional instantons break it down to $\mathbb{Z}_2$, or should we just ignore the fractional instantons and think the $\mathbb{Z}_6$ remains a good symmetry? Neither.

In fact the global symmetries that are anomalous only in the fractional instanton backgrounds are \textit{not} unaffected by the anomaly, nor are they fully broken. Rather such symmetries can be recovered in the theory on $\mathbb{R}^4$ as `non-invertible chiral symmetries' \cite{Cordova:2022ieu,Choi:2022jqy}, which must act both on local operators and on 't Hooft lines as well. This is possible in our theory because of the nontrivial quotient, which means that the theory has $\mathbb{Z}_3$ magnetic representations, and these 't Hooft lines enjoy a magnetic one-form symmetry in the IR. Then the actual structure of the PQ symmetry after accounting for explicit breaking is\footnote{Note that we classify the `non-invertible symmetry' a bit differently than e.g. our friends in \cite{Cordova:2023her}. Whenever there is a non-invertible symmetry, it can be stacked on top of any invertible symmetry to create another non-invertible symmetry. So in their classification we have $\mathbb{Z}_2 \ (\text{invertible}) \subset \mathbb{Z}_6 \ (\text{non-invertible})$, and it is technically true that we could create $\mathbb{Z}_6$ worth of non-invertible symmetry defect operators. We describe it differently here because the important point from a model-building perspective is that there is a $\mathbb{Z}_3$ subgroup which is protected \textit{only} by non-invertible symmetry and not any invertible symmetry, and it is this non-invertible symmetry which is automatically broken when going to a UV theory which introduces magnetic monopoles.} 
\begin{equation}
    U(1)_{\rm PQ} \rightarrow \mathbb{Z}_2 \ (\text{invertible}) \times \mathbb{Z}_3 \ (\text{non-invertible}).
\end{equation}
Furthermore, as effective field theorists we have learned non-invertible symmetries can offer a useful guide for ultraviolet model building \cite{Cordova:2022fhg,Cordova:2024ypu,Cordova:2023her}: When we find an interesting spurion for a non-invertible symmetry in a low-energy theory, we should look for a UV theory which breaks the low-energy $\mathbb{Z}_3^{(1)}$ magnetic one-form symmetry by upgrading these IR 't Hooft lines to dynamical magnetic monopoles, and so upgrades the relevant instantons to existing on $\mathbb{R}^4$. 
Embedding the quark flavor symmetry in the UV theory of $SU(9)$ quark color-flavor unification provides non-invertible symmetry breaking, and so can generate $\Delta(\text{PQ})=2$ operators.\footnote{Parenthetically we remark that this same feature means our UV embedding does not have a domain wall problem. 
Further study of the use of non-invertible symmetry to solve the domain wall problem will appear soon in the context of the DFSZ theory \cite{Choi:2024a}.}

\section{\texorpdfstring{$SU(9)$}{SU(9)} Theory and Instanton NDA} \label{sec:su9}

As in \cite{Cordova:2024ypu}, our understanding of non-invertible symmetry breaking impels us to consider UV completing the $SU(3)^2/\mathbb{Z}_3$ theory to the color-flavor unified $SU(9)$ theory in Table \ref{tab:uv_charges}. The yukawa interactions have a familiar form
\begin{equation}
    \mathcal{L} \supset y_t \Phi_1 Q \bar u + y_b \tilde{\Phi}_2 Q \bar d,
\end{equation}
where the yukawas are now just two numbers and we have omitted the color-flavor indices contracted between the left-handed and right-handed quarks. The difference in the sizes of these couplings remains the only source of irreducible $\mathbb{Z}_2$ breaking in this theory, since the $SU(9)$ gluons treat up and down symmetrically. 
The $SU(9)$ Yang-Mills will be broken spontaneously down to the $SU(3)^2/\mathbb{Z}_3$ theory considered above by the three-index symmetric scalar $\Xi^{ABC}$,
\begin{equation}
    \langle \Xi^{ABC} \rangle = v_9 \epsilon^{abc} \epsilon^{\alpha \beta \gamma},
\end{equation}
where e.g. $A=1..9$ is an $SU(9)$ index which can be thought of as a multi-index in terms of the $SU(3)_C$ index $a=1..3$ and the $SU(3)_H$ index $\alpha=1..3$ as $A=11,12,..,23,33$.
In the expanded scalar potential, $\Xi$ may couple to the two Higgs doublets as $\lambda |\Xi^{ABC}|^2(|\Phi_1|^2 + |\Phi_2|^2)$ and is not a source of irreducible $\mathbb{Z}_2$ breaking.

A complete model will require also scalar fields to spontaneously break $SU(3)_H$ and generate the non-trivial flavor and CP structure of the SM yukawas, but these issues are separate from our main points here. A proof of principle was given in \cite{Cordova:2024ypu}, and studying more realistic such structures remains an interesting and necessary direction for connecting color-flavor unification to the real world. We note that eventually there does need to be some further $\mathbb{Z}_2$ symmetry-breaking because the flavor structures in the up and down sectors do not coincide, but presumably this need not be a larger effect than $y_t\gg y_b$, which we have already said we can ignore.

\begin{table}\centering
\large
\renewcommand{\arraystretch}{1.3}
\begin{tabular}{|c|c|c|c|c|}  \hline
 & $SU(9)$ & $SU(2)_L$ & $U(1)_Y$ & $U(1)_{\rm PQ}$ \\ \hline

${Q}$ & $9$ & $2$ & $+1$ & $0$  \\ \hline

${\bar u}$ & $\bar 9$ & $-$ & $-4$ & $1$ \\ \hline

${\bar d}$ & $\bar 9$ & $-$ & $+2$ & $1$ \\ \hline \hline

$\Phi_1$ & $-$& $2$ & $+3$ & $-1$ \\ \hline

$\Phi_2$ & $-$& $2$ & $+3$ & $1$ \\ \hline

$\Xi$ & $165$ & $-$ & $0$ & $0$ \\ \hline

\end{tabular}\caption{Symmetry and matter content of the $SU(9)$ theory including the SM quarks, the 2HDs, and the three-index symmetric scalar $\Xi$ whose condensation breaks $SU(9) \rightarrow SU(3)^2/\mathbb{Z}_3$.}\label{tab:uv_charges}
\end{table}

The only feature of this UV embedding of the $SU(3)^2/\mathbb{Z}_3$ which we need to make use of is the non-invertible symmetry breaking by $SU(9)$ instantons, whose main effect is to generate $m_{12}^2$ and softly break the PQ symmetry in the scalar potential. 
The 't Hooft vertices \cite{tHooft:1976snw} are simple enough to understand in this case, but we can make use of the general technology of Csaki, Tito D'Agnolo, Kuflik, Ruhdorfer \cite{Csaki:2023ziz} to systematically estimate the size of the mass parameter generated by $SU(9)$ color-flavor instantons. Alternatively see also \cite{Sesma:2024tcd} for a functional approach.

\subsection{Generation of \texorpdfstring{$m_{12}^2$}{m_{12}^2}}

\begin{figure}
    \centering
    \includegraphics[trim={0 0.5cm 0 0},clip,width=0.5\linewidth]{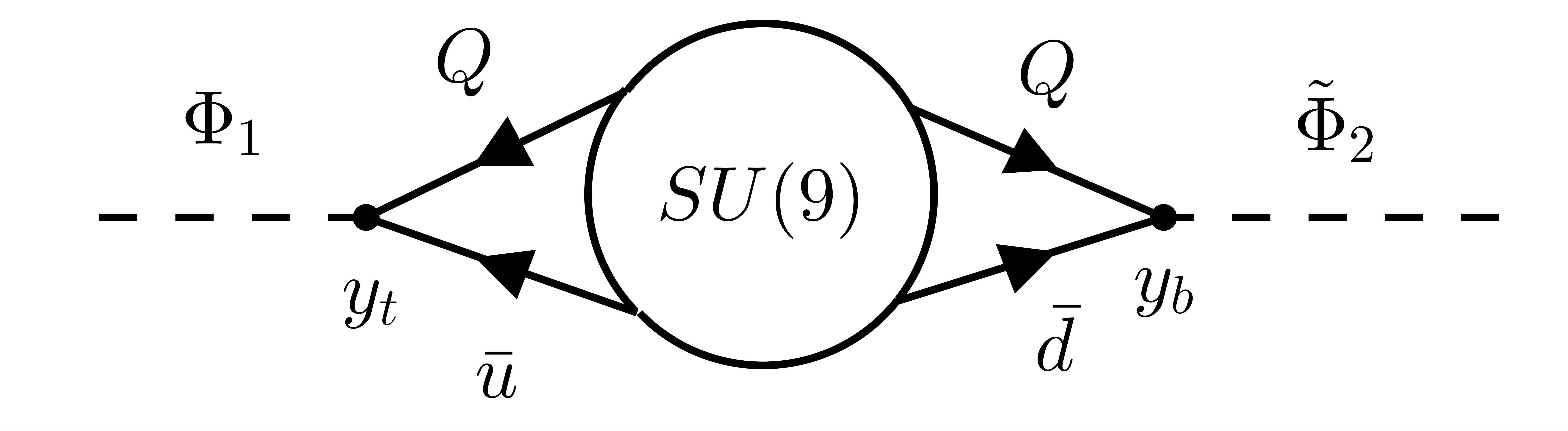}
    \caption{The 1-instanton 't Hooft vertex of the $SU(9)$ 2HDM with both sets of fermion lines closed off with flavor-symmetric yukawas to produce $m_{12}^2$.}
    \label{fig:2HDMinst}
\end{figure}

In the UV $SU(9)$ theory with RG-invariant scale $\Lambda_9$ which is Higgsed at the scale $v_9$, instanton NDA gives us (see Figure \ref{fig:2HDMinst})
\begin{equation}
    m_{12}^2 \sim y_t y_b C_9 \left( \frac{8 \pi^2}{g(v_9)^2}\right)^{2\cdot 9} \int \frac{d\rho}{\rho^5} (\Lambda_9 \rho)^{b_0} e^{- 2\pi^2 \rho^2 v_9^2} \rho^2,
\end{equation}
\noindent where $b_0$ is the leading UV beta function coefficient and $C_9$ is an instanton density factor computed by 't Hooft which depends also on the multiplicities of charged fields.\footnote{We note that this often-quoted size of the exponential suppression from Higgsing $\exp(- 2\pi^2 \rho^2 v_9^2)$ is in fact the result from Affleck \cite{Affleck:1980mp} for $SU(2)$ broken by the fundamental. In more general scenarios one finds an order-one representation-theoretic factor $q$, which depends on the details of the breaking $SU(9)\rightarrow SU(3)^2/\mathbb{Z}_3$ or $\rightarrow (SU(3)_C\times U(1)_H)/\mathbb{Z}_3$ or $\rightarrow SU(3)_C$. Absorbing this into $\tilde{v}_9 \equiv v_9/\sqrt{q}$ and accounting for the implicit dependence of $\Lambda_9$ on $v_9$, one finds this scales the result as $q^{1-b_0/2}$. We thank Gongjun Choi for calling our attention to this.} This integral over instanton size $\rho$ gives
\begin{equation}
    m_{12}^2 \sim y_t y_b v_9^2 \frac{C_9 \pi^2}{(\sqrt{2} \pi)^{b_0}} \left( \frac{8 \pi^2}{g(v_9)^2}\right)^{18} e^{-\frac{8\pi^2}{g(v_9)^2}}  \Gamma\left(\frac{b_0}{2}-1\right).
\end{equation}

As to the beta function, in general 
$b_0 = \frac{11}{3} N - \frac{4}{3} n_f I_f - \frac{1}{3} n_s I_s - \frac{1}{6} n_r I_r$ for charged Dirac fermions, complex scalars, and real scalars respectively with Dynkin indices $I_i$.
In the $SU(9)$ phase the SM fermions are simply contained in 2 fundamental Dirac fermions with index $\half$, and we Higgs this with one three-index symmetric complex scalar $\Xi$ with index $33$, for a coefficient of $b_0=62/3$.

\begin{figure}
    \centering
    \includegraphics[width=0.5\linewidth]{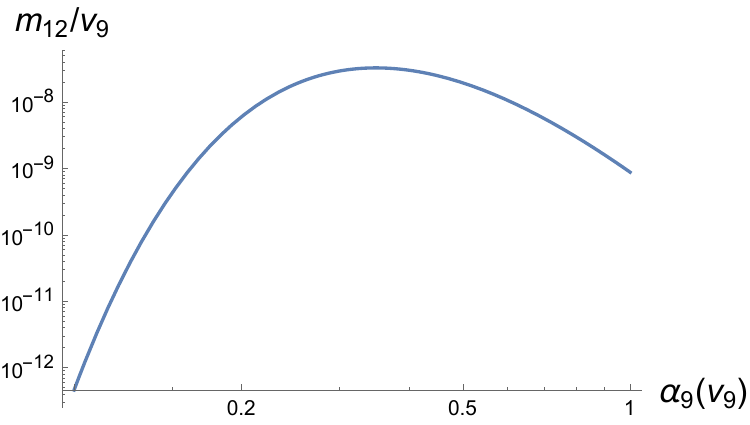}
    \caption{The mass parameter of the 2HDM generated as soft breaking of the Peccei-Quinn symmetry by the color-flavor unified instantons as a function of $\alpha(v_9)$ in our toy UV model.}
    \label{fig:massgen}
\end{figure}

With this matter content we plot $m_{12}/v_9$ as a function of the $SU(9)$ gauge coupling $\alpha_9(v_9) \equiv g(v_9)^2/(4\pi)$ in Figure \ref{fig:massgen}. Note that due to the non-trivial embedding $\pi_3\left(SU(9)/\left[SU(3)^2/\mathbb{Z}_3\right]\right)=\mathbb{Z}_3$, this coupling jumps by a factor of three across $v_9$, so $\alpha_9(v_9) = 3 \alpha_s(v_9)$ where $\alpha_s$ is the strong coupling of $SU(3)_C$ \cite{Csaki:1998vv}.
We remark that there is considerable freedom to alter the sizes of these effects---for one because additional matter in the UV will change the beta function $b_0$, and for a smaller beta function the suppression of $m_{12}/v_9$ will become less severe. More starkly, the $SU(3)_C$ coupling can easily grow between the scale $v_3$ where $SU(3)_H$ is broken and the scale $v_9$, for example if some of the colored degrees of freedom in $\Xi$ get masses moderately below $v_9$ (see \cite{Cordova:2024ypu}).

Henceforth we treat $m_{12}^2$ as a free parameter which is the spurion for $U(1)_{\rm PQ}$-breaking. 
The structure of the couplings between the Higgs scalars and the fermions means that the PQ violation of the 't Hooft vertex is transmitted to the scalar potential solely into $m_{12}^2$ at lowest order, so that the breaking in the scalar sector is effectively soft to one-loop. This means the inclusion of $m_{12}^2\neq 0$ is the leading effect in our discussion of where we end up in the 2HDM parameter space in Section \ref{sec:paramspace}.

\subsection{Suppressed PQ-violating Quartics}

\begin{figure}
    \centering
    \includegraphics[trim={0 0.5cm 0 0},clip,width=0.5\linewidth]{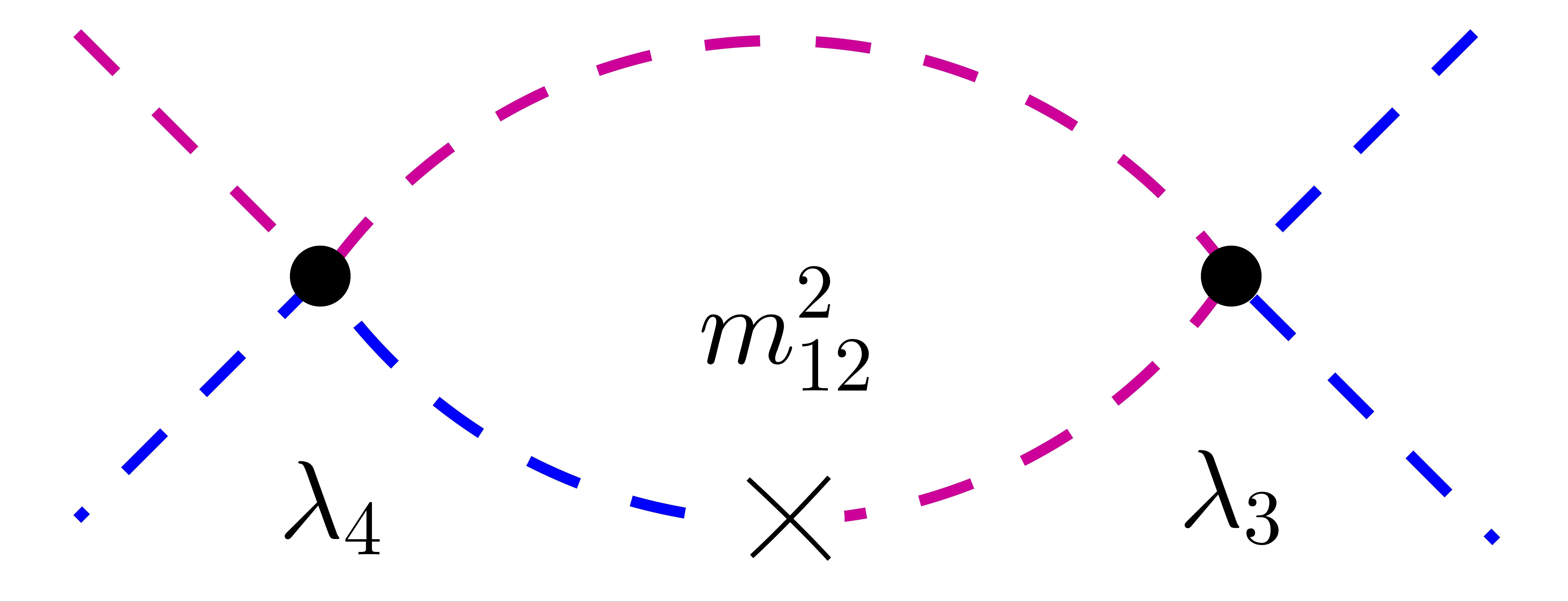}
    \caption{One diagram contributing to the generation of $\lambda_6$ at one loop, where $\Phi_1$ lines are colored blue and $\Phi_2$ lines purple. The PQ-violating spurion $m_{12}^2$ necessarily appears.}
    \label{fig:Quartic1loop1}
\end{figure}

Since the PQ-breaking in the scalar sector is soft, it does not lead to any divergent diagrams generating PQ-violating quartic operators upon renormalization group evolution, as diagrams like Figure \ref{fig:Quartic1loop1} are finite. The only generation of these quartics will take place upon integrating out the `heavy' Higgses---to the extent that there is any splitting between the scalar masses, which in our target parameter space is quite slight in any case. 

At higher-loop order the protection of effective softness in the scalar sector will break down, as ultimately the four-fermi 't Hooft vertex is an irrelevant interaction. As our purpose in this work is not to write down a complete model, we will not take pains to compute these loop diagrams. Neither the finite pieces of the one-loop diagrams nor the renormalization from higher-loop diagrams skews us away from the model with softly-broken PQ enough to affect near-term viability of this model.

\section{2HDM Parameter Space} \label{sec:paramspace}

In this section we consider where in the 2HDM parameter space we have landed from the non-invertible Peccei-Quinn breaking.
As discussed above, the violation of the scalar sector symmetries in the dimensionless couplings is loop suppressed and small. Then to quite good approximation we are in the parameter space where the PQ symmetry is only broken by the relevant $m_{12}^2$, and the $\mathbb{Z}_2$ symmetry is only broken by the relevant $m_{11}^2 > m_{22}^2$. This fits in a simple part of the Exceptional Region of Parameter Space ERPS4 \cite{Haber:2021zva}, which makes accessible the alignment without decoupling limit as Haber \& Silva discuss.

In particular, at lowest order our scenario bears out the softly-broken ERPS4 region of Sec VI of \cite{Haber:2021zva}. Defining $\lambda_1 = \lambda_2 \equiv \lambda$ and $R = (\lambda_3+\lambda_4)/\lambda$, the spectrum is simply
\begin{align}
    m_A^2 &= m_{12}^2 \frac{2}{s_{2\beta}} \\ 
    m_{H^{\pm}}^2 &= m_A^2 - \half \lambda_4 v^2 \\
    m_{h,H}^2 &= \half \left[m_A^2 + \lambda v^2 \pm \sqrt{\left[m_A^2 - \lambda v^2 \left(c_{2\beta}^2 + R s_{2\beta}^2 \right)\right]^2+\lambda^2 s_{2\beta}^2 c_{2\beta}^2 (1-R)^2 v^4}\right] \\
    c_{2\beta} &= \frac{m_{22}^2 - m_{11}^2}{m_{22}^2 + m_{11}^2 +\lambda v^2}, 
\end{align}
where $c_{2\beta} \equiv \cos(2\beta), s_{2\beta} \equiv \sin(2\beta)$. How close we are to the alignment limit is parametrized by 
\begin{equation}
    \cos(\beta - \alpha) = \frac{\lambda v^2 s_{2\beta} c_{2\beta} (1-R)}{2\sqrt{(m_H^2 - m_h^2)\left[m_H^2 - \lambda v^2 (1 - \half s_{2\beta}^2 (1-R))\right]}},
\end{equation}
with exact alignment achieved for $\cos(\beta - \alpha) = 0$. This is exhibited in e.g. the yukawa couplings to $h$ which are modified proportionally to the SM with the factors
\begin{align}
    \kappa_u &= \sin \left( \beta - \alpha \right) + \cot \beta \cos \left( \beta - \alpha \right) \\
    \kappa_{d,\ell} &= \sin \left( \beta - \alpha \right) - \tan \beta \cos \left( \beta - \alpha \right),
\end{align}
where $y_f^{\rm 2HDM} = \kappa_f y_f^{\rm SM}$.

In a general 2HDM the pseudoscalar mass $m_A$ is bounded from below by roughly $m_h/2$ having not observed $h \rightarrow AA$ decays at the LHC. 
As recently reviewed in \cite{Karan:2023xze,Karan:2024kgr}, the current bound on how close we are to the alignment limit is $\cos(\beta-\alpha)\lesssim.05$ and we require additionally $\tan\beta \gtrsim 5$. In order to satisfy those bounds we have to deviate from the exact $\mathbb{Z}_2$ limit with $m_{11}=m_{22}$ to get large $\tan\beta$, which is anyway an obvious expectation given $y_t \gg y_b$.  

As we are restricted to a small subspace of the type II 2HDM parameter space by the softly-broken $U(1)_{\rm PQ} \times \mathbb{Z}_2$ symmetry, satisfying these requirements will restrict us to $m_A \gtrsim 120 \text{ GeV}$. While not as light as the pseudoscalar can be in general, this is still well within the alignment without decoupling regime. Generally, the values of $(\tan \beta, \lambda, R, m_A)$ which we can reach are tightly correlated. To the extent we are at $\infty \gg \tan \beta \gg 1$, as required by Higgs coupling measurements, good alignment will require $R\sim 1$. This implies 
\begin{align}
 m_{h,H}^2 &= \half \left[m_A^2 + \lambda v^2 \pm \sqrt{\left[m_A^2 - \lambda v^2 \left(c_{2\beta}^2 + R s_{2\beta}^2 \right)\right]^2+\lambda^2 s_{2\beta}^2 c_{2\beta}^2 (1-R)^2 v^4}\right] \nonumber\\
 \cos(\beta - \alpha) \sim 0 &\Rightarrow \half \bigg[ m_A^2 + \lambda v^2 \pm \left|m_A^2 - \lambda v^2 \left(c_{2\beta}^2 + R s_{2\beta}^2 \right)\right|\bigg] \nonumber\\
 R\sim 1 &\Rightarrow \half \bigg[ m_A^2 + \lambda v^2 \pm \left|m_A^2 - \lambda v^2\right|\bigg] \nonumber\\
 m_{h,H}^2 &\sim \half \lambda v^2, m_A^2,
\end{align}
which means that we require $\lambda \sim 2 \lambda_{\rm SM}$ and necessarily find $m_H^2 \sim m_A^2$.

\section{\texorpdfstring{$A = a$}{A=a}}\label{sec:A=a}

Any 2HDM contains a scalar field charged under a $U(1)$ Peccei-Quinn symmetry which condenses, and empirically we know the infrared PQ symmetry must be approximate such that the would-be Goldstone mode becomes a massive pseudo-Goldstone. \textit{If} the only explicit breaking of such a PQ symmetry is by instantons, then very generally such a setup bears out an axion solution to strong CP. In the minimal infrared setup of Weinberg \& Wilczek, which merely extends the SM by a second Higgs doublet, the only source of breaking is instantons of $SU(3)_C$, but this leads to a light axion of mass $m_a \sim \Lambda_{\rm QCD}^2/v_{\rm EW}$ with large couplings to SM fermions. So this \textit{visible} axion was very quickly ruled out. 

Our focus thus far has been the natural realization of the 2HDM alignment limit as a result of a UV Peccei-Quinn symmetry which we were led to by recognizing an IR non-invertible PQ symmetry in a theory of gauged quark flavor.
But this has lead us also to simple unified UV theories of the SM fermions in which additional small instantons contribute to the explicit violation of the PQ symmetry. In such a UV theory, the pseudoscalar pseudo-Goldstone axion mass can be dominated by the effects of these small instantons, taking on a much larger value $m_a \sim m_{12}^2/v_{\rm EW}$. Such a `heavy QCD axion' can still solve the strong CP problem, and in fact has some comparative virtues.

We note that there have been a variety of constructions of heavy QCD axions, e.g. in models which enlarge QCD with another confining group \cite{Dimopoulos:1979pp,Tye:1981zy,Holdom:1982ex,Flynn:1987rs,Gherghetta:2016fhp,Gaillard:2018xgk,Valenti:2022tsc}, in models which `deconstruct' QCD \cite{Agrawal:2017ksf,Csaki:2019vte}, in models utilizing discrete symmetries which relate our QCD to another \cite{Rubakov:1997vp,Berezhiani:2000gh,Hook:2014cda,Fukuda:2015ana,Albaid:2015axa,Dimopoulos:2016lvn,Chiang:2016eav,Hook:2019qoh,Dunsky:2023ucb}, or in models of five-dimensional dynamics or compositeness \cite{Gherghetta:2020keg,Gherghetta:2020ofz,Gupta:2020vxb,Gherghetta:2021jnn,Aoki:2024usv}.
In general, heavier QCD axions have various interesting probes \cite{Kelly:2020dda,Bertholet:2021hjl,Chakraborty:2021wda,Kivel:2022emq,Dunsky:2022uoq,Co:2022bqq,Bedi:2022qrd,Ahmadvand:2023lpp,Bedi:2024wqg,Chakraborty:2024tyx}. These are great ideas which, it is fair to say, all involve a fair amount of extra structure.

In this revival of the visible Weiberg-Wilczek axion in color-flavor unification, the axion lives \textit{entirely} in the two Higgses which provide masses to the Standard Model fermions, so it is maximally tied to the structure of the SM. Furthermore, since the small instantons generating the heavy axion mass exist from the gauging of some of the approximate global symmetries of the SM, we have no need to introduce new fermions.

\subsection{Peccei-Quinn Weinberg-Wilczek Axion}

Let us recall which mode is the axion, $A$. The scalar sector has $U(1)$ currents
\begin{equation}
    J_Y^\mu = 3 \left(\Phi_1^\dagger \partial^\mu \Phi_1 + \Phi_2^\dagger \partial^\mu \Phi_2\right), \qquad J_{\rm PQ}^\mu = - \Phi_1^\dagger \partial^\mu \Phi_1 + \Phi_2^\dagger \partial^\mu \Phi_2.
\end{equation}
For the PQWW axion, the identification of the pseudoscalar modes is trivial: The $\mathbb{Z}_2$-even combination is eaten by the $Z$-boson, and the axion should be identified with the orthogonal $\mathbb{Z}_2$-odd linear combination of the neutral pseudoscalar modes of $\Phi_1$ and $\Phi_2$. Being the \textit{visible} axion, this pseudoscalar has the large Yukawa couplings to the SM fermions familiar from the 2HDM
\begin{equation}
    \mathcal{L} \supset i \varsigma_f y_f^{\rm SM} A \bar \psi_f \gamma_5 \psi_f, 
\end{equation}
where $y_f^{\rm SM}$ is the SM yukawa and $\varsigma_{u} = \cot \beta, \varsigma_{d,\ell}=-\tan\beta$. 

In a model of the invisible axion, there is a huge separation between the PQ-breaking scale  and the mass of the axion, $f_a\gg m_a\sim \Lambda_{\rm QCD}$, and so below $f_a$ it is sensible to integrate out the radial mode of the singlet scalar PQ-breaking field. Then in the effective theory below $f_a$, the axion non-linearly realizes the PQ symmetry. However, in the visible axion model the scale of Peccei-Quinn breaking is given by the electroweak scale $v_{EW}$. 
Then there is no `separation of scales' between the axion decay constant and the axion mass within which we may formulate an effective field theory where the PQ symmetry is non-linearly realized by the 2HDM pseudoscalar. 

\subsection{No Quality Problem} \label{sec:quality}
Recall that the invisible axion has a severe quality problem: We expect to have no control over global-symmetry-violating operators induced by Planck scale physics, and must consider the effects of 
\begin{equation}
    \mathcal{L} \supset a_n M_{\rm pl}^{n-4} \phi^n,
\end{equation}
where $\phi$ is the complex scalar singlet which Higgses PQ $\langle \phi \rangle = f_a$ whose phase is the invisible axion. This leads to an axion potential like 
\begin{equation}
    V(a) \sim \Lambda_{\rm QCD}^4 \left(1 - \cos (a)\right) + f_a^4 \left(\frac{f_a}{M_{\rm pl}}\right)^{n-4} \left(1 - \cos (n a + \varphi_n)\right),
\end{equation}
where $\varphi_n$ is the phase of $a_n$. Observational constraints from stellar cooling have pushed invisible axion models up to the regime $f_a \gtrsim 10^8 \text{ GeV }\gg \Lambda_{\rm QCD}$, so these latter contributions threaten to destabilize the axion solution by localizing it away from $\bar\theta_{\rm eff} = 0$---unless the couplings $a_n$ or their phases are very very nongeneric.

In contrast, in the 2HDM alignment without decoupling limit, the spontaneous PQ breaking occurs at the electroweak scale. And in our revival of the visible axion using small instantons, the instanton induced potential is at that same scale as well. So then what is the status of the quality problem? Which putative quantum-gravity-generated global-symmetry-violating operators are of worry?

The effects of such an operator should vanish in the limit $M_{\rm pl} \rightarrow \infty$, which is responsible for the folklore that we should only consider irrelevant operators---at least if we don't have any other scales around. If all we had was the 2HDM, then we should consider gravity inducing the following operator of greatest concern
\begin{equation}
    \mathcal{L} \supset a_6 \left( \Phi_1^\dagger \Phi_2\right)^3/M_{\rm pl}^2
\end{equation}
which leads to a visible axion potential of
\begin{equation}
    V(a) \sim m_{12}^4 \left(1 - \cos (2a)\right) + v_{\rm EW}^4 \left(\frac{v_{\rm EW}}{M_{\rm pl}}\right)^{2} \left(1 - \cos (6 a + \varphi_6)\right),
\end{equation}
and the theory is automatically safe from these quality concerns---already the first irrelevant operator is suppressed enough that the mechanism is robust to even $\mathcal{O}(1)$ Planck-scale global symmetry violation.

However, we have introduced additional UV dynamics here which can make the concern slightly more tangible. In particular we can consider the operator 
\begin{equation}
    \mathcal{L} \supset a_2 \left( \Phi_1^\dagger \Phi_2\right) |\Xi|^4 /M_{\rm pl}^2,
\end{equation}
utilizing the $SU(9)$-breaking scalar $\Xi$, which leads to a visible axion potential of
\begin{equation}
    V(a) \sim m_{12}^4 \left(1 - \cos (2a)\right) + v_{\rm EW}^2 v_9^2 \left(\frac{v_{9}}{M_{\rm pl}}\right)^{2} \left(1 - \cos (2 a + \varphi_2)\right).
\end{equation}
If $\varphi_2$ is a generic phase, then we must demand the latter term perturbs the minimization $a \sim 0$ by less than the observed constraint $\bar \theta \lesssim 10^{-10}$. This term is enhanced over the first by $v_9^2/v_{\rm EW}^2$ so this enhancement must be drowned out by $v_9^2/M_{\rm Pl}^2$
\begin{equation}
    v_9^2/M_{\rm Pl}^2 \lesssim 10^{-10} v_{\rm EW}^2/v_9^2 \ \leftrightarrow \ v_9 \lesssim 10^8 \text{ GeV}
\end{equation}

This operator is not a concern to the extent that the color-flavor breaking scale $v_9$ can be far below the Planck scale.\footnote{We note that there can also be quality concerns from Planck-suppressed dimension-6 SMEFT operators that violate CP, but the resulting bounds on $v_9/M_{\rm pl}$ are less stringent than the above in this case, see \cite{Bedi:2022qrd}.} The scale of gauged quark flavor must be far above collider scales to avoid constraints on flavor-changing neutral currents \cite{EuropeanStrategyforParticlePhysicsPreparatoryGroup:2019qin}. The exact constraint depends upon the details of its breaking, but $v_3 \gtrsim 10^6 \text{ GeV}$ is conservative. Remarkably, there is no concern of Peccei-Quinn quality if the flavor-breaking scale is within striking distance in the mid term, as was seen also in \cite{Cordova:2024ypu}.

\section{Conclusion}

Already the application of generalized symmetries to model building has guided us to interesting nonperturbative effects in models of non-trivial flavor unification. 
In the quark sector of the 1HDM the identification of a non-invertible Peccei-Quinn symmetry pointed to a simple UV revival of the massless up quark solution \cite{Cordova:2024ypu}. The massless quark solution solves strong CP using a PQ symmetry which is \textit{not} spontaneously broken, and is the solution which is closest to utilizing solely the dynamics of the SM fields. 

The possibility of solving strong CP using a spontaneously broken symmetry requires an additional scalar to provide the axion as a goldstone mode. The simplest such possibility was described by Weinberg and Wilczek and has long been ruled out. 
Here again a non-invertible symmetry analysis in the IR theory with another Higgs doublet has now pointed us to a simple UV revival of the Weinberg-Wilczek visible axion, which has built into it a natural way to get alignment without decoupling in the 2HDM and so have this new pseudoscalar live at accessible scales. Elsewhere some of us will report on interesting features from utilizing insights into non-invertible Peccei-Quinn symmetries in the invisible DFSZ axion model \cite{Choi:2024a}.

The theory we have written down is not yet fully realistic, and it remains an important problem to write convincing models of spontaneous breaking of color-flavor unification which generate the observed yukawas, for which as yet there is only a proof of principle \cite{Cordova:2024ypu}. It would be interesting also to consider the early universe cosmology of this model and the dynamics of its topological defects. More broadly, models of minimal non-trivial gauge-flavor unification of the Standard Model fermions merit further attention \cite{Davighi:2022qgb,Davighi:2022fer,Davighi:2022vpl,Davighi:2023xqn,Cordova:2022fhg,Cordova:2024ypu}. 

\section*{Acknowledgements}

We thank Brian Batell for collaboration during the early stages of this project. We are grateful to Sungwoo Hong and Gongjun Choi for comments on a draft of this manuscript, and SK thanks them also for related discussions and collaboration.
This work is partially supported by the National Science Foundation under grant PHY-2412701. 

\bibliographystyle{jhep}
\bibliography{nipq2hdm}

\providecommand{\href}[2]{#2}\begingroup\begin{thebibliography}{100}

\bibitem{Gaiotto:2014kfa}
D.~Gaiotto, A.~Kapustin, N.~Seiberg and B.~Willett, \emph{{Generalized Global Symmetries}}, \href{http://dx.doi.org/10.1007/JHEP02(2015)172}{\emph{JHEP} {\bf 02} (2015) 172} [\href{https://arxiv.org/abs/1412.5148}{{\tt arXiv:1412.5148}}].

\bibitem{Brennan:2023mmt}
T.~D. Brennan and S.~Hong, \emph{{Introduction to Generalized Global Symmetries in QFT and Particle Physics}},  \href{https://arxiv.org/abs/2306.00912}{{\tt arXiv:2306.00912}}.

\bibitem{Schafer-Nameki:2023jdn}
S.~Schafer-Nameki, \emph{{ICTP lectures on (non-)invertible generalized symmetries}}, \href{http://dx.doi.org/10.1016/j.physrep.2024.01.007}{\emph{Phys. Rept.} {\bf 1063} (2024) 1--55} [\href{https://arxiv.org/abs/2305.18296}{{\tt arXiv:2305.18296}}].

\bibitem{Shao:2023gho}
S.-H. Shao, \emph{{What's Done Cannot Be Undone: TASI Lectures on Non-Invertible Symmetry}},  \href{https://arxiv.org/abs/2308.00747}{{\tt arXiv:2308.00747}}.

\bibitem{Cordova:2022ieu}
C.~Cordova and K.~Ohmori, \emph{{Noninvertible Chiral Symmetry and Exponential Hierarchies}}, \href{http://dx.doi.org/10.1103/PhysRevX.13.011034}{\emph{Phys. Rev. X} {\bf 13} (2023) 011034} [\href{https://arxiv.org/abs/2205.06243}{{\tt arXiv:2205.06243}}].

\bibitem{Choi:2022jqy}
Y.~Choi, H.~T. Lam and S.-H. Shao, \emph{{Noninvertible Global Symmetries in the Standard Model}}, \href{http://dx.doi.org/10.1103/PhysRevLett.129.161601}{\emph{Phys. Rev. Lett.} {\bf 129} (2022) 161601} [\href{https://arxiv.org/abs/2205.05086}{{\tt arXiv:2205.05086}}].

\bibitem{Cordova:2018cvg}
C.~C\'{o}rdova, T.~T. Dumitrescu and K.~Intriligator, \emph{{Exploring 2-Group Global Symmetries}}, \href{http://dx.doi.org/10.1007/JHEP02(2019)184}{\emph{JHEP} {\bf 02} (2019) 184} [\href{https://arxiv.org/abs/1802.04790}{{\tt arXiv:1802.04790}}].

\bibitem{Benini:2018reh}
F.~Benini, C.~C\'ordova and P.-S. Hsin, \emph{{On 2-Group Global Symmetries and their Anomalies}}, \href{http://dx.doi.org/10.1007/JHEP03(2019)118}{\emph{JHEP} {\bf 03} (2019) 118} [\href{https://arxiv.org/abs/1803.09336}{{\tt arXiv:1803.09336}}].

\bibitem{Cordova:2022qtz}
C.~C\'{o}rdova and S.~Koren, \emph{{Higher Flavor Symmetries in the Standard Model}}, \href{http://dx.doi.org/10.1002/andp.202300031}{\emph{Annalen Phys.} {\bf 535} (2023) 2300031} [\href{https://arxiv.org/abs/2212.13193}{{\tt arXiv:2212.13193}}].

\bibitem{Koren:2024xof}
S.~Koren and A.~Martin, \emph{{Fractionally Charged Particles at the Energy Frontier: The SM Gauge Group and One-Form Global Symmetry}},  \href{https://arxiv.org/abs/2406.17850}{{\tt arXiv:2406.17850}}.

\bibitem{Cordova:2022fhg}
C.~Cordova, S.~Hong, S.~Koren and K.~Ohmori, \emph{{Neutrino Masses from Generalized Symmetry Breaking}}, \href{http://dx.doi.org/10.1103/PhysRevX.14.031033}{\emph{Phys. Rev. X} {\bf 14} (2024) 031033} [\href{https://arxiv.org/abs/2211.07639}{{\tt arXiv:2211.07639}}].

\bibitem{Cordova:2024ypu}
C.~Cordova, S.~Hong and S.~Koren, \emph{{Non-Invertible Peccei-Quinn Symmetry and the Massless Quark Solution to the Strong CP Problem}},  \href{https://arxiv.org/abs/2402.12453}{{\tt arXiv:2402.12453}}.

\bibitem{Brennan:2020ehu}
T.~D. Brennan and C.~C\'{o}rdova, \emph{{Axions, higher-groups, and emergent symmetry}}, \href{http://dx.doi.org/10.1007/JHEP02(2022)145}{\emph{JHEP} {\bf 02} (2022) 145} [\href{https://arxiv.org/abs/2011.09600}{{\tt arXiv:2011.09600}}].

\bibitem{Hidaka:2020izy}
Y.~Hidaka, M.~Nitta and R.~Yokokura, \emph{{Global 3-group symmetry and 't Hooft anomalies in axion electrodynamics}}, \href{http://dx.doi.org/10.1007/JHEP01(2021)173}{\emph{JHEP} {\bf 01} (2021) 173} [\href{https://arxiv.org/abs/2009.14368}{{\tt arXiv:2009.14368}}].

\bibitem{Hidaka:2021mml}
Y.~Hidaka, M.~Nitta and R.~Yokokura, \emph{{Topological axion electrodynamics and 4-group symmetry}}, \href{http://dx.doi.org/10.1016/j.physletb.2021.136762}{\emph{Phys. Lett. B} {\bf 823} (2021) 136762} [\href{https://arxiv.org/abs/2107.08753}{{\tt arXiv:2107.08753}}].

\bibitem{Hidaka:2021kkf}
Y.~Hidaka, M.~Nitta and R.~Yokokura, \emph{{Global 4-group symmetry and \textquoteright{}t Hooft anomalies in topological axion electrodynamics}}, \href{http://dx.doi.org/10.1093/ptep/ptab150}{\emph{PTEP} {\bf 2022} (2022) 04A109} [\href{https://arxiv.org/abs/2108.12564}{{\tt arXiv:2108.12564}}].

\bibitem{Nakajima:2022feg}
T.~Nakajima, T.~Sakai and R.~Yokokura, \emph{{Higher-group structure in 2n-dimensional axion-electrodynamics}}, \href{http://dx.doi.org/10.1007/JHEP01(2023)150}{\emph{JHEP} {\bf 01} (2023) 150} [\href{https://arxiv.org/abs/2211.13861}{{\tt arXiv:2211.13861}}].

\bibitem{Brennan:2023kpw}
T.~D. Brennan, S.~Hong and L.-T. Wang, \emph{{Coupling a Cosmic String to a TQFT}}, \href{http://dx.doi.org/10.1007/JHEP03(2024)145}{\emph{JHEP} {\bf 03} (2024) 145} [\href{https://arxiv.org/abs/2302.00777}{{\tt arXiv:2302.00777}}].

\bibitem{Anber:2024gis}
M.~M. Anber and S.~Y.~L. Chan, \emph{{Global aspects of 3-form gauge theory: implications for axion-Yang-Mills systems}}, \href{http://dx.doi.org/10.1007/JHEP10(2024)113}{\emph{JHEP} {\bf 10} (2024) 113} [\href{https://arxiv.org/abs/2407.03416}{{\tt arXiv:2407.03416}}].

\bibitem{Yokokura:2022alv}
R.~Yokokura, \emph{{Non-invertible symmetries in axion electrodynamics}},  \href{https://arxiv.org/abs/2212.05001}{{\tt arXiv:2212.05001}}.

\bibitem{Choi:2022fgx}
Y.~Choi, H.~T. Lam and S.-H. Shao, \emph{{Non-invertible Gauss law and axions}}, \href{http://dx.doi.org/10.1007/JHEP09(2023)067}{\emph{JHEP} {\bf 09} (2023) 067} [\href{https://arxiv.org/abs/2212.04499}{{\tt arXiv:2212.04499}}].

\bibitem{Hidaka:2024kfx}
Y.~Hidaka, M.~Nitta and R.~Yokokura, \emph{{Selection rules of topological solitons from non-invertible symmetries in axion electrodynamics}},  \href{https://arxiv.org/abs/2411.05434}{{\tt arXiv:2411.05434}}.

\bibitem{DelZotto:2024ngj}
M.~Del~Zotto, M.~Dell'Acqua and E.~Riedel~G\r{a}rding, \emph{{The Higher Structure of Symmetries of Axion-Maxwell Theory}},  \href{https://arxiv.org/abs/2411.09685}{{\tt arXiv:2411.09685}}.

\bibitem{Cordova:2023ent}
C.~Cordova and K.~Ohmori, \emph{{Quantum duality in electromagnetism and the fine structure constant}}, \href{http://dx.doi.org/10.1103/PhysRevD.109.105019}{\emph{Phys. Rev. D} {\bf 109} (2024) 105019} [\href{https://arxiv.org/abs/2307.12927}{{\tt arXiv:2307.12927}}].

\bibitem{Reece:2023iqn}
M.~Reece, \emph{{Axion-gauge coupling quantization with a twist}}, \href{http://dx.doi.org/10.1007/JHEP10(2023)116}{\emph{JHEP} {\bf 10} (2023) 116} [\href{https://arxiv.org/abs/2309.03939}{{\tt arXiv:2309.03939}}].

\bibitem{Choi:2023pdp}
Y.~Choi, M.~Forslund, H.~T. Lam and S.-H. Shao, \emph{{Quantization of Axion-Gauge Couplings and Noninvertible Higher Symmetries}}, \href{http://dx.doi.org/10.1103/PhysRevLett.132.121601}{\emph{Phys. Rev. Lett.} {\bf 132} (2024) 121601} [\href{https://arxiv.org/abs/2309.03937}{{\tt arXiv:2309.03937}}].

\bibitem{Cordova:2023her}
C.~Cordova, S.~Hong and L.-T. Wang, \emph{{Axion domain walls, small instantons, and non-invertible symmetry breaking}}, \href{http://dx.doi.org/10.1007/JHEP05(2024)325}{\emph{JHEP} {\bf 05} (2024) 325} [\href{https://arxiv.org/abs/2309.05636}{{\tt arXiv:2309.05636}}].

\bibitem{Reece:2024wrn}
M.~Reece, \emph{{Extra-Dimensional Axion Expectations}},  \href{https://arxiv.org/abs/2406.08543}{{\tt arXiv:2406.08543}}.

\bibitem{Craig:2024dnl}
N.~Craig and M.~Kongsore, \emph{{High-Quality Axions from Higher-Form Symmetries in Extra Dimensions}},  \href{https://arxiv.org/abs/2408.10295}{{\tt arXiv:2408.10295}}.

\bibitem{Aloni:2024jpb}
D.~Aloni, E.~Garc\'\i{}a-Valdecasas, M.~Reece and M.~Suzuki, \emph{{Spontaneously broken (-1)-form U(1) symmetries}}, \href{http://dx.doi.org/10.21468/SciPostPhys.17.2.031}{\emph{SciPost Phys.} {\bf 17} (2024) 031} [\href{https://arxiv.org/abs/2402.00117}{{\tt arXiv:2402.00117}}].

\bibitem{Chen:2024tsx}
S.~Chen, A.~Cherman, G.~Choi and M.~Neuzil, \emph{{Cheshire $\theta$ terms, Aharonov-Bohm effects, and axions}},  \href{https://arxiv.org/abs/2410.23355}{{\tt arXiv:2410.23355}}.

\bibitem{Cordova:2022rer}
C.~C\'{o}rdova, K.~Ohmori and T.~Rudelius, \emph{{Generalized Symmetry Breaking Scales and Weak Gravity Conjectures}},  \href{https://arxiv.org/abs/2202.05866}{{\tt arXiv:2202.05866}}.

\bibitem{Brennan:2022tyl}
T.~D. Brennan, C.~Cordova and T.~T. Dumitrescu, \emph{{Line Defect Quantum Numbers \& Anomalies}},  \href{https://arxiv.org/abs/2206.15401}{{\tt arXiv:2206.15401}}.

\bibitem{Brennan:2023tae}
T.~D. Brennan, \emph{{A New Solution to the Callan Rubakov Effect}},  \href{https://arxiv.org/abs/2309.00680}{{\tt arXiv:2309.00680}}.

\bibitem{Brennan:2024iau}
T.~D. Brennan, J.~S. Grewal and E.~Y. Yang, \emph{{Revisiting Scattering Enhancement from the Aharonov-Bohm Effect}},  \href{https://arxiv.org/abs/2411.10526}{{\tt arXiv:2411.10526}}.

\bibitem{vanBeest:2023dbu}
M.~van Beest, P.~Boyle~Smith, D.~Delmastro, Z.~Komargodski and D.~Tong, \emph{{Monopoles, Scattering, and Generalized Symmetries}},  \href{https://arxiv.org/abs/2306.07318}{{\tt arXiv:2306.07318}}.

\bibitem{vanBeest:2023mbs}
M.~van Beest, P.~Boyle~Smith, D.~Delmastro, R.~Mouland and D.~Tong, \emph{{Fermion-monopole scattering in the Standard Model}}, \href{http://dx.doi.org/10.1007/JHEP08(2024)004}{\emph{JHEP} {\bf 08} (2024) 004} [\href{https://arxiv.org/abs/2312.17746}{{\tt arXiv:2312.17746}}].

\bibitem{Davighi:2019rcd}
J.~Davighi, B.~Gripaios and N.~Lohitsiri, \emph{{Global anomalies in the Standard Model(s) and Beyond}}, \href{http://dx.doi.org/10.1007/JHEP07(2020)232}{\emph{JHEP} {\bf 07} (2020) 232} [\href{https://arxiv.org/abs/1910.11277}{{\tt arXiv:1910.11277}}].

\bibitem{Davighi:2020kok}
J.~Davighi and N.~Lohitsiri, \emph{{Omega vs. pi, and 6d anomaly cancellation}}, \href{http://dx.doi.org/10.1007/JHEP05(2021)267}{\emph{JHEP} {\bf 05} (2021) 267} [\href{https://arxiv.org/abs/2012.11693}{{\tt arXiv:2012.11693}}].

\bibitem{Davighi:2024zip}
J.~Davighi, A.~Greljo and N.~Selimovic, \emph{{Topological Portal to the Dark Sector}},  \href{https://arxiv.org/abs/2401.09528}{{\tt arXiv:2401.09528}}.

\bibitem{Davighi:2024zjp}
J.~Davighi and N.~Lohitsiri, \emph{{WZW terms without anomalies: generalised symmetries in chiral Lagrangians}},  \href{https://arxiv.org/abs/2407.20340}{{\tt arXiv:2407.20340}}.

\bibitem{Anber:2021upc}
M.~M. Anber and E.~Poppitz, \emph{{Nonperturbative effects in the Standard Model with gauged 1-form symmetry}}, \href{http://dx.doi.org/10.1007/JHEP12(2021)055}{\emph{JHEP} {\bf 12} (2021) 055} [\href{https://arxiv.org/abs/2110.02981}{{\tt arXiv:2110.02981}}].

\bibitem{Cheung:2024ypq}
C.~Cheung, M.~Derda, J.-H. Kim, V.~Nevoa, I.~Rothstein and N.~Shah, \emph{{Generalized symmetry in dynamical gravity}}, \href{http://dx.doi.org/10.1007/JHEP10(2024)007}{\emph{JHEP} {\bf 10} (2024) 007} [\href{https://arxiv.org/abs/2403.01837}{{\tt arXiv:2403.01837}}].

\bibitem{Kobayashi:2024yqq}
T.~Kobayashi and H.~Otsuka, \emph{{Non-invertible flavor symmetries in magnetized extra dimensions}},  \href{https://arxiv.org/abs/2408.13984}{{\tt arXiv:2408.13984}}.

\bibitem{Kobayashi:2024cvp}
T.~Kobayashi, H.~Otsuka and M.~Tanimoto, \emph{{Yukawa textures from non-invertible symmetries}},  \href{https://arxiv.org/abs/2409.05270}{{\tt arXiv:2409.05270}}.

\bibitem{Funakoshi:2024uvy}
S.~Funakoshi, T.~Kobayashi and H.~Otsuka, \emph{{Quantum aspects of non-invertible flavor symmetries in intersecting/magnetized D-brane models}},  \href{https://arxiv.org/abs/2412.12524}{{\tt arXiv:2412.12524}}.

\bibitem{Cherman:2022eml}
A.~Cherman, T.~Jacobson and M.~Neuzil, \emph{{1-form symmetry versus large N QCD}},  \href{https://arxiv.org/abs/2209.00027}{{\tt arXiv:2209.00027}}.

\bibitem{Cherman:2023xok}
A.~Cherman and T.~Jacobson, \emph{{Emergent 1-form symmetries}}, \href{http://dx.doi.org/10.1103/PhysRevD.109.125013}{\emph{Phys. Rev. D} {\bf 109} (2024) 125013} [\href{https://arxiv.org/abs/2304.13751}{{\tt arXiv:2304.13751}}].

\bibitem{Hinterbichler:2022agn}
K.~Hinterbichler, D.~M. Hofman, A.~Joyce and G.~Mathys, \emph{{Gravity as a gapless phase and biform symmetries}}, \href{http://dx.doi.org/10.1007/JHEP02(2023)151}{\emph{JHEP} {\bf 02} (2023) 151} [\href{https://arxiv.org/abs/2205.12272}{{\tt arXiv:2205.12272}}].

\bibitem{Hinterbichler:2024cxn}
K.~Hinterbichler, A.~Joyce and G.~Mathys, \emph{{Impossible symmetries and conformal gravity}}, \href{http://dx.doi.org/10.1103/PhysRevD.110.085003}{\emph{Phys. Rev. D} {\bf 110} (2024) 085003} [\href{https://arxiv.org/abs/2403.03256}{{\tt arXiv:2403.03256}}].

\bibitem{Choi:2022rfe}
Y.~Choi, H.~T. Lam and S.-H. Shao, \emph{{Noninvertible Time-Reversal Symmetry}}, \href{http://dx.doi.org/10.1103/PhysRevLett.130.131602}{\emph{Phys. Rev. Lett.} {\bf 130} (2023) 131602} [\href{https://arxiv.org/abs/2208.04331}{{\tt arXiv:2208.04331}}].

\bibitem{Arbalestrier:2024oqg}
A.~Arbalestrier, R.~Argurio and L.~Tizzano, \emph{{Noninvertible axial symmetry in QED comes full circle}}, \href{http://dx.doi.org/10.1103/PhysRevD.110.105012}{\emph{Phys. Rev. D} {\bf 110} (2024) 105012} [\href{https://arxiv.org/abs/2405.06596}{{\tt arXiv:2405.06596}}].

\bibitem{Heidenreich:2020pkc}
B.~Heidenreich, J.~McNamara, M.~Montero, M.~Reece, T.~Rudelius and I.~Valenzuela, \emph{{Chern-Weil global symmetries and how quantum gravity avoids them}}, \href{http://dx.doi.org/10.1007/JHEP11(2021)053}{\emph{JHEP} {\bf 11} (2021) 053} [\href{https://arxiv.org/abs/2012.00009}{{\tt arXiv:2012.00009}}].

\bibitem{Heidenreich:2021xpr}
B.~Heidenreich, J.~McNamara, M.~Montero, M.~Reece, T.~Rudelius and I.~Valenzuela, \emph{{Non-invertible global symmetries and completeness of the spectrum}}, \href{http://dx.doi.org/10.1007/JHEP09(2021)203}{\emph{JHEP} {\bf 09} (2021) 203} [\href{https://arxiv.org/abs/2104.07036}{{\tt arXiv:2104.07036}}].

\bibitem{Heidenreich:2023pbi}
B.~Heidenreich, J.~McNamara and M.~Reece, \emph{{Non-standard axion electrodynamics and the dual Witten effect}}, \href{http://dx.doi.org/10.1007/JHEP01(2024)120}{\emph{JHEP} {\bf 01} (2024) 120} [\href{https://arxiv.org/abs/2309.07951}{{\tt arXiv:2309.07951}}].

\bibitem{Fan:2021ntg}
J.~Fan, K.~Fraser, M.~Reece and J.~Stout, \emph{{Axion Mass from Magnetic Monopole Loops}}, \href{http://dx.doi.org/10.1103/PhysRevLett.127.131602}{\emph{Phys. Rev. Lett.} {\bf 127} (2021) 131602} [\href{https://arxiv.org/abs/2105.09950}{{\tt arXiv:2105.09950}}].

\bibitem{McNamara:2022lrw}
J.~McNamara and M.~Reece, \emph{{Reflections on Parity Breaking}},  \href{https://arxiv.org/abs/2212.00039}{{\tt arXiv:2212.00039}}.

\bibitem{Asadi:2022vys}
P.~Asadi, S.~Homiller, Q.~Lu and M.~Reece, \emph{{Chiral Nelson-Barr models: Quality and cosmology}}, \href{http://dx.doi.org/10.1103/PhysRevD.107.115012}{\emph{Phys. Rev. D} {\bf 107} (2023) 115012} [\href{https://arxiv.org/abs/2212.03882}{{\tt arXiv:2212.03882}}].

\bibitem{Garcia-Valdecasas:2024cqn}
E.~Garc\'\i{}a-Valdecasas, M.~Reece and M.~Suzuki, \emph{{Monopole Breaking of Chern-Weil Symmetries}},  \href{https://arxiv.org/abs/2408.00067}{{\tt arXiv:2408.00067}}.

\bibitem{Wan:2019gqr}
Z.~Wan and J.~Wang, \emph{{Beyond Standard Models and Grand Unifications: Anomalies, Topological Terms, and Dynamical Constraints via Cobordisms}}, \href{http://dx.doi.org/10.1007/JHEP07(2020)062}{\emph{JHEP} {\bf 07} (2020) 062} [\href{https://arxiv.org/abs/1910.14668}{{\tt arXiv:1910.14668}}].

\bibitem{Wang:2020xyo}
J.~Wang, \emph{{Anomaly and Cobordism Constraints Beyond the Standard Model: Topological Force}},  \href{https://arxiv.org/abs/2006.16996}{{\tt arXiv:2006.16996}}.

\bibitem{Wang:2021ayd}
J.~Wang, Z.~Wan and Y.-Z. You, \emph{{Cobordism and deformation class of the standard model}}, \href{http://dx.doi.org/10.1103/PhysRevD.106.L041701}{\emph{Phys. Rev. D} {\bf 106} (2022) L041701} [\href{https://arxiv.org/abs/2112.14765}{{\tt arXiv:2112.14765}}].

\bibitem{Wang:2021vki}
J.~Wang and Y.-Z. You, \emph{{Gauge Enhanced Quantum Criticality Between Grand Unifications: Categorical Higher Symmetry Retraction}},  \href{https://arxiv.org/abs/2111.10369}{{\tt arXiv:2111.10369}}.

\bibitem{Wang:2021hob}
J.~Wang and Y.-Z. You, \emph{{Gauge enhanced quantum criticality beyond the standard model}}, \href{http://dx.doi.org/10.1103/PhysRevD.106.025013}{\emph{Phys. Rev. D} {\bf 106} (2022) 025013} [\href{https://arxiv.org/abs/2106.16248}{{\tt arXiv:2106.16248}}].

\bibitem{Wang:2022eag}
J.~Wang, Z.~Wan and Y.-Z. You, \emph{{Proton stability: From the standard model to beyond grand unification}}, \href{http://dx.doi.org/10.1103/PhysRevD.106.025016}{\emph{Phys. Rev. D} {\bf 106} (2022) 025016} [\href{https://arxiv.org/abs/2204.08393}{{\tt arXiv:2204.08393}}].

\bibitem{Putrov:2023jqi}
P.~Putrov and J.~Wang, \emph{{Categorical Symmetry of the Standard Model from Gravitational Anomaly}},  \href{https://arxiv.org/abs/2302.14862}{{\tt arXiv:2302.14862}}.

\bibitem{Wang:2023tbj}
J.~Wang, \emph{{Family Puzzle, Framing Topology, $c_-=24$ and 3(E8)$_1$ Conformal Field Theories: 48/16 = 45/15 = 24/8 =3}},  \href{https://arxiv.org/abs/2312.14928}{{\tt arXiv:2312.14928}}.

\bibitem{Peccei:1977hh}
R.~D. Peccei and H.~R. Quinn, \emph{{CP Conservation in the Presence of Instantons}}, \href{http://dx.doi.org/10.1103/PhysRevLett.38.1440}{\emph{Phys. Rev. Lett.} {\bf 38} (1977) 1440--1443}.

\bibitem{Peccei:1977ur}
R.~D. Peccei and H.~R. Quinn, \emph{{Constraints Imposed by CP Conservation in the Presence of Instantons}}, \href{http://dx.doi.org/10.1103/PhysRevD.16.1791}{\emph{Phys. Rev. D} {\bf 16} (1977) 1791--1797}.

\bibitem{Weinberg:1977ma}
S.~Weinberg, \emph{{A New Light Boson?}}, \href{http://dx.doi.org/10.1103/PhysRevLett.40.223}{\emph{Phys. Rev. Lett.} {\bf 40} (1978) 223--226}.

\bibitem{Wilczek:1977pj}
F.~Wilczek, \emph{{Problem of Strong $P$ and $T$ Invariance in the Presence of Instantons}}, \href{http://dx.doi.org/10.1103/PhysRevLett.40.279}{\emph{Phys. Rev. Lett.} {\bf 40} (1978) 279--282}.

\bibitem{Davidson:2005cw}
S.~Davidson and H.~E. Haber, \emph{{Basis-independent methods for the two-Higgs-doublet model}}, \href{http://dx.doi.org/10.1103/PhysRevD.72.099902}{\emph{Phys. Rev. D} {\bf 72} (2005) 035004} [\href{https://arxiv.org/abs/hep-ph/0504050}{{\tt arXiv:hep-ph/0504050}}].

\bibitem{Ferreira:2009wh}
P.~M. Ferreira, H.~E. Haber and J.~P. Silva, \emph{{Generalized CP symmetries and special regions of parameter space in the two-Higgs-doublet model}}, \href{http://dx.doi.org/10.1103/PhysRevD.79.116004}{\emph{Phys. Rev. D} {\bf 79} (2009) 116004} [\href{https://arxiv.org/abs/0902.1537}{{\tt arXiv:0902.1537}}].

\bibitem{Haber:2020wco}
H.~E. Haber, \emph{{A tale of three diagonalizations}}, \href{http://dx.doi.org/10.1142/S0217751X21300027}{\emph{Int. J. Mod. Phys. A} {\bf 36} (2021) 2130003} [\href{https://arxiv.org/abs/2009.03990}{{\tt arXiv:2009.03990}}].

\bibitem{Draper:2020tyq}
P.~Draper, A.~Ekstedt and H.~E. Haber, \emph{{A natural mechanism for approximate Higgs alignment in the 2HDM}}, \href{http://dx.doi.org/10.1007/JHEP05(2021)235}{\emph{JHEP} {\bf 05} (2021) 235} [\href{https://arxiv.org/abs/2011.13159}{{\tt arXiv:2011.13159}}].

\bibitem{Gunion:2002zf}
J.~F. Gunion and H.~E. Haber, \emph{{The CP conserving two Higgs doublet model: The Approach to the decoupling limit}}, \href{http://dx.doi.org/10.1103/PhysRevD.67.075019}{\emph{Phys. Rev. D} {\bf 67} (2003) 075019} [\href{https://arxiv.org/abs/hep-ph/0207010}{{\tt arXiv:hep-ph/0207010}}].

\bibitem{Branco:2011iw}
G.~C. Branco, P.~M. Ferreira, L.~Lavoura, M.~N. Rebelo, M.~Sher and J.~P. Silva, \emph{{Theory and phenomenology of two-Higgs-doublet models}}, \href{http://dx.doi.org/10.1016/j.physrep.2012.02.002}{\emph{Phys. Rept.} {\bf 516} (2012) 1--102} [\href{https://arxiv.org/abs/1106.0034}{{\tt arXiv:1106.0034}}].

\bibitem{Haber:2021zva}
H.~E. Haber and J.~P. Silva, \emph{{Exceptional regions of the 2HDM parameter space}}, \href{http://dx.doi.org/10.1103/PhysRevD.103.115012}{\emph{Phys. Rev. D} {\bf 103} (2021) 115012} [\href{https://arxiv.org/abs/2102.07136}{{\tt arXiv:2102.07136}}].

\bibitem{Tong:2017oea}
D.~Tong, \emph{{Line Operators in the Standard Model}}, \href{http://dx.doi.org/10.1007/JHEP07(2017)104}{\emph{JHEP} {\bf 07} (2017) 104} [\href{https://arxiv.org/abs/1705.01853}{{\tt arXiv:1705.01853}}].

\bibitem{Anber:2021iip}
M.~M. Anber, S.~Hong and M.~Son, \emph{{New anomalies, TQFTs, and confinement in bosonic chiral gauge theories}}, \href{http://dx.doi.org/10.1007/JHEP02(2022)062}{\emph{JHEP} {\bf 02} (2022) 062} [\href{https://arxiv.org/abs/2109.03245}{{\tt arXiv:2109.03245}}].

\bibitem{Li:2024nuo}
H.-L. Li and L.-X. Xu, \emph{{Understanding the SM gauge group from SMEFT}}, \href{http://dx.doi.org/10.1007/JHEP07(2024)199}{\emph{JHEP} {\bf 07} (2024) 199} [\href{https://arxiv.org/abs/2404.04229}{{\tt arXiv:2404.04229}}].

\bibitem{Alonso:2024pmq}
R.~Alonso, D.~Dimakou and M.~West, \emph{{Fractional-charge hadrons and leptons to tell the Standard Model group apart}},  \href{https://arxiv.org/abs/2404.03438}{{\tt arXiv:2404.03438}}.

\bibitem{Hsin:2024lya}
P.-S. Hsin and J.~Gomis, \emph{{Detecting Standard Model Gauge Group from Generalized Fractional Quantum Hall Effect}},  \href{https://arxiv.org/abs/2411.18160}{{\tt arXiv:2411.18160}}.

\bibitem{Dierigl:2024cxm}
M.~Dierigl and D.~Novi\v{c}i\'c, \emph{{The axion is going dark}}, \href{http://dx.doi.org/10.1007/JHEP12(2024)104}{\emph{JHEP} {\bf 12} (2024) 104} [\href{https://arxiv.org/abs/2409.02180}{{\tt arXiv:2409.02180}}].

\bibitem{Choi:2024a}
G.~Choi, S.~Hong and S.~Koren, \emph{{Non-Invertible Peccei-Quinn Symmetry Breaking Solves the DFSZ Domain Wall Problem}},  \href{https://arxiv.org/abs/In preparation}{{\tt arXiv:In preparation}}.

\bibitem{tHooft:1976snw}
G.~'t~Hooft, \emph{{Computation of the Quantum Effects Due to a Four-Dimensional Pseudoparticle}}, \href{http://dx.doi.org/10.1103/PhysRevD.14.3432}{\emph{Phys. Rev. D} {\bf 14} (1976) 3432--3450}.

\bibitem{Csaki:2023ziz}
C.~Cs\'aki, R.~T. D'Agnolo, E.~Kuflik and M.~Ruhdorfer, \emph{{Instanton NDA and applications to axion models}}, \href{http://dx.doi.org/10.1007/JHEP04(2024)074}{\emph{JHEP} {\bf 04} (2024) 074} [\href{https://arxiv.org/abs/2311.09285}{{\tt arXiv:2311.09285}}].

\bibitem{Sesma:2024tcd}
P.~Sesma, \emph{{A functional treatment of small instanton-induced axion potentials}},  \href{https://arxiv.org/abs/2411.00101}{{\tt arXiv:2411.00101}}.

\bibitem{Affleck:1980mp}
I.~Affleck, \emph{{On Constrained Instantons}}, \href{http://dx.doi.org/10.1016/0550-3213(81)90307-2}{\emph{Nucl. Phys. B} {\bf 191} (1981) 429}.

\bibitem{Csaki:1998vv}
C.~Csaki and H.~Murayama, \emph{{Instantons in partially broken gauge groups}}, \href{http://dx.doi.org/10.1016/S0550-3213(98)00448-9}{\emph{Nucl. Phys. B} {\bf 532} (1998) 498--526} [\href{https://arxiv.org/abs/hep-th/9804061}{{\tt arXiv:hep-th/9804061}}].

\bibitem{Karan:2023xze}
A.~Karan, V.~Miralles and A.~Pich, \emph{{Aligned two Higgs doublet model and the global fits}}, \href{http://dx.doi.org/10.22323/1.449.0053}{\emph{PoS} {\bf EPS-HEP2023} (2024) 053} [\href{https://arxiv.org/abs/2312.00514}{{\tt arXiv:2312.00514}}].

\bibitem{Karan:2024kgr}
A.~Karan, A.~M. Coutinho, V.~Miralles and A.~Pich, \emph{{Status of the Aligned Two Higgs Doublet Model in the low mass region}},  in \emph{{42nd International Conference on High Energy Physics}}, 9 2024.
\newblock [\href{https://arxiv.org/abs/2409.14934}{{\tt arXiv:2409.14934}}].

\bibitem{Dimopoulos:1979pp}
S.~Dimopoulos, \emph{{A Solution of the Strong {CP} Problem in Models With Scalars}}, \href{http://dx.doi.org/10.1016/0370-2693(79)91233-4}{\emph{Phys. Lett. B} {\bf 84} (1979) 435--439}.

\bibitem{Tye:1981zy}
S.~H.~H. Tye, \emph{{A Superstrong Force With a Heavy Axion}}, \href{http://dx.doi.org/10.1103/PhysRevLett.47.1035}{\emph{Phys. Rev. Lett.} {\bf 47} (1981) 1035}.

\bibitem{Holdom:1982ex}
B.~Holdom and M.~E. Peskin, \emph{{Raising the Axion Mass}}, \href{http://dx.doi.org/10.1016/0550-3213(82)90228-0}{\emph{Nucl. Phys. B} {\bf 208} (1982) 397--412}.

\bibitem{Flynn:1987rs}
J.~M. Flynn and L.~Randall, \emph{{A Computation of the Small Instanton Contribution to the Axion Potential}}, \href{http://dx.doi.org/10.1016/0550-3213(87)90089-7}{\emph{Nucl. Phys. B} {\bf 293} (1987) 731--739}.

\bibitem{Gherghetta:2016fhp}
T.~Gherghetta, N.~Nagata and M.~Shifman, \emph{{A Visible QCD Axion from an Enlarged Color Group}}, \href{http://dx.doi.org/10.1103/PhysRevD.93.115010}{\emph{Phys. Rev. D} {\bf 93} (2016) 115010} [\href{https://arxiv.org/abs/1604.01127}{{\tt arXiv:1604.01127}}].

\bibitem{Gaillard:2018xgk}
M.~K. Gaillard, M.~B. Gavela, R.~Houtz, P.~Quilez and R.~Del~Rey, \emph{{Color unified dynamical axion}}, \href{http://dx.doi.org/10.1140/epjc/s10052-018-6396-6}{\emph{Eur. Phys. J. C} {\bf 78} (2018) 972} [\href{https://arxiv.org/abs/1805.06465}{{\tt arXiv:1805.06465}}].

\bibitem{Valenti:2022tsc}
A.~Valenti, L.~Vecchi and L.-X. Xu, \emph{{Grand Color axion}}, \href{http://dx.doi.org/10.1007/JHEP10(2022)025}{\emph{JHEP} {\bf 10} (2022) 025} [\href{https://arxiv.org/abs/2206.04077}{{\tt arXiv:2206.04077}}].

\bibitem{Agrawal:2017ksf}
P.~Agrawal and K.~Howe, \emph{{Factoring the Strong CP Problem}}, \href{http://dx.doi.org/10.1007/JHEP12(2018)029}{\emph{JHEP} {\bf 12} (2018) 029} [\href{https://arxiv.org/abs/1710.04213}{{\tt arXiv:1710.04213}}].

\bibitem{Csaki:2019vte}
C.~Cs\'aki, M.~Ruhdorfer and Y.~Shirman, \emph{{UV Sensitivity of the Axion Mass from Instantons in Partially Broken Gauge Groups}}, \href{http://dx.doi.org/10.1007/JHEP04(2020)031}{\emph{JHEP} {\bf 04} (2020) 031} [\href{https://arxiv.org/abs/1912.02197}{{\tt arXiv:1912.02197}}].

\bibitem{Rubakov:1997vp}
V.~A. Rubakov, \emph{{Grand unification and heavy axion}}, \href{http://dx.doi.org/10.1134/1.567390}{\emph{JETP Lett.} {\bf 65} (1997) 621--624} [\href{https://arxiv.org/abs/hep-ph/9703409}{{\tt arXiv:hep-ph/9703409}}].

\bibitem{Berezhiani:2000gh}
Z.~Berezhiani, L.~Gianfagna and M.~Giannotti, \emph{{Strong CP problem and mirror world: The Weinberg-Wilczek axion revisited}}, \href{http://dx.doi.org/10.1016/S0370-2693(00)01392-7}{\emph{Phys. Lett. B} {\bf 500} (2001) 286--296} [\href{https://arxiv.org/abs/hep-ph/0009290}{{\tt arXiv:hep-ph/0009290}}].

\bibitem{Hook:2014cda}
A.~Hook, \emph{{Anomalous solutions to the strong CP problem}}, \href{http://dx.doi.org/10.1103/PhysRevLett.114.141801}{\emph{Phys. Rev. Lett.} {\bf 114} (2015) 141801} [\href{https://arxiv.org/abs/1411.3325}{{\tt arXiv:1411.3325}}].

\bibitem{Fukuda:2015ana}
H.~Fukuda, K.~Harigaya, M.~Ibe and T.~T. Yanagida, \emph{{Model of visible QCD axion}}, \href{http://dx.doi.org/10.1103/PhysRevD.92.015021}{\emph{Phys. Rev. D} {\bf 92} (2015) 015021} [\href{https://arxiv.org/abs/1504.06084}{{\tt arXiv:1504.06084}}].

\bibitem{Albaid:2015axa}
A.~Albaid, M.~Dine and P.~Draper, \emph{{Strong CP and SUZ$_{2}$}}, \href{http://dx.doi.org/10.1007/JHEP12(2015)046}{\emph{JHEP} {\bf 12} (2015) 046} [\href{https://arxiv.org/abs/1510.03392}{{\tt arXiv:1510.03392}}].

\bibitem{Dimopoulos:2016lvn}
S.~Dimopoulos, A.~Hook, J.~Huang and G.~Marques-Tavares, \emph{{A collider observable QCD axion}}, \href{http://dx.doi.org/10.1007/JHEP11(2016)052}{\emph{JHEP} {\bf 11} (2016) 052} [\href{https://arxiv.org/abs/1606.03097}{{\tt arXiv:1606.03097}}].

\bibitem{Chiang:2016eav}
C.-W. Chiang, H.~Fukuda, M.~Ibe and T.~T. Yanagida, \emph{{750 GeV diphoton resonance in a visible heavy QCD axion model}}, \href{http://dx.doi.org/10.1103/PhysRevD.93.095016}{\emph{Phys. Rev. D} {\bf 93} (2016) 095016} [\href{https://arxiv.org/abs/1602.07909}{{\tt arXiv:1602.07909}}].

\bibitem{Hook:2019qoh}
A.~Hook, S.~Kumar, Z.~Liu and R.~Sundrum, \emph{{High Quality QCD Axion and the LHC}}, \href{http://dx.doi.org/10.1103/PhysRevLett.124.221801}{\emph{Phys. Rev. Lett.} {\bf 124} (2020) 221801} [\href{https://arxiv.org/abs/1911.12364}{{\tt arXiv:1911.12364}}].

\bibitem{Dunsky:2023ucb}
D.~I. Dunsky, L.~J. Hall and K.~Harigaya, \emph{{A heavy QCD axion and the mirror world}}, \href{http://dx.doi.org/10.1007/JHEP02(2024)212}{\emph{JHEP} {\bf 02} (2024) 212} [\href{https://arxiv.org/abs/2302.04274}{{\tt arXiv:2302.04274}}].

\bibitem{Gherghetta:2020keg}
T.~Gherghetta, V.~V. Khoze, A.~Pomarol and Y.~Shirman, \emph{{The Axion Mass from 5D Small Instantons}}, \href{http://dx.doi.org/10.1007/JHEP03(2020)063}{\emph{JHEP} {\bf 03} (2020) 063} [\href{https://arxiv.org/abs/2001.05610}{{\tt arXiv:2001.05610}}].

\bibitem{Gherghetta:2020ofz}
T.~Gherghetta and M.~D. Nguyen, \emph{{A Composite Higgs with a Heavy Composite Axion}}, \href{http://dx.doi.org/10.1007/JHEP12(2020)094}{\emph{JHEP} {\bf 12} (2020) 094} [\href{https://arxiv.org/abs/2007.10875}{{\tt arXiv:2007.10875}}].

\bibitem{Gupta:2020vxb}
R.~S. Gupta, V.~V. Khoze and M.~Spannowsky, \emph{{Small instantons and the strong CP problem in composite Higgs models}}, \href{http://dx.doi.org/10.1103/PhysRevD.104.075011}{\emph{Phys. Rev. D} {\bf 104} (2021) 075011} [\href{https://arxiv.org/abs/2012.00017}{{\tt arXiv:2012.00017}}].

\bibitem{Gherghetta:2021jnn}
T.~Gherghetta and A.~Pomarol, \emph{{Small instantons in weakly-gauged holographic models}}, \href{http://dx.doi.org/10.1007/JHEP11(2021)136}{\emph{JHEP} {\bf 11} (2021) 136} [\href{https://arxiv.org/abs/2110.01762}{{\tt arXiv:2110.01762}}].

\bibitem{Aoki:2024usv}
T.~Aoki, M.~Ibe, S.~Shirai and K.~Watanabe, \emph{{Small instanton effects on composite axion mass}}, \href{http://dx.doi.org/10.1007/JHEP07(2024)269}{\emph{JHEP} {\bf 07} (2024) 269} [\href{https://arxiv.org/abs/2404.19342}{{\tt arXiv:2404.19342}}].

\bibitem{Kelly:2020dda}
K.~J. Kelly, S.~Kumar and Z.~Liu, \emph{{Heavy axion opportunities at the DUNE near detector}}, \href{http://dx.doi.org/10.1103/PhysRevD.103.095002}{\emph{Phys. Rev. D} {\bf 103} (2021) 095002} [\href{https://arxiv.org/abs/2011.05995}{{\tt arXiv:2011.05995}}].

\bibitem{Bertholet:2021hjl}
E.~Bertholet, S.~Chakraborty, V.~Loladze, T.~Okui, A.~Soffer and K.~Tobioka, \emph{{Heavy QCD axion at Belle II: Displaced and prompt signals}}, \href{http://dx.doi.org/10.1103/PhysRevD.105.L071701}{\emph{Phys. Rev. D} {\bf 105} (2022) L071701} [\href{https://arxiv.org/abs/2108.10331}{{\tt arXiv:2108.10331}}].

\bibitem{Chakraborty:2021wda}
S.~Chakraborty, M.~Kraus, V.~Loladze, T.~Okui and K.~Tobioka, \emph{{Heavy QCD axion in b\textrightarrow{}s transition: Enhanced limits and projections}}, \href{http://dx.doi.org/10.1103/PhysRevD.104.055036}{\emph{Phys. Rev. D} {\bf 104} (2021) 055036} [\href{https://arxiv.org/abs/2102.04474}{{\tt arXiv:2102.04474}}].

\bibitem{Kivel:2022emq}
A.~Kivel, J.~Laux and F.~Yu, \emph{{Supersizing axions with small size instantons}}, \href{http://dx.doi.org/10.1007/JHEP11(2022)088}{\emph{JHEP} {\bf 11} (2022) 088} [\href{https://arxiv.org/abs/2207.08740}{{\tt arXiv:2207.08740}}].

\bibitem{Dunsky:2022uoq}
D.~I. Dunsky, L.~J. Hall and K.~Harigaya, \emph{{Dark Radiation Constraints on Heavy QCD Axions}}, \href{http://dx.doi.org/10.1007/JHEP04(2024)130}{\emph{JHEP} {\bf 04} (2024) 130} [\href{https://arxiv.org/abs/2205.11540}{{\tt arXiv:2205.11540}}].

\bibitem{Co:2022bqq}
R.~T. Co, S.~Kumar and Z.~Liu, \emph{{Searches for heavy QCD axions via dimuon final states}}, \href{http://dx.doi.org/10.1007/JHEP02(2023)111}{\emph{JHEP} {\bf 02} (2023) 111} [\href{https://arxiv.org/abs/2210.02462}{{\tt arXiv:2210.02462}}].

\bibitem{Bedi:2022qrd}
R.~S. Bedi, T.~Gherghetta and M.~Pospelov, \emph{{Enhanced EDMs from small instantons}}, \href{http://dx.doi.org/10.1103/PhysRevD.106.015030}{\emph{Phys. Rev. D} {\bf 106} (2022) 015030} [\href{https://arxiv.org/abs/2205.07948}{{\tt arXiv:2205.07948}}].

\bibitem{Ahmadvand:2023lpp}
M.~Ahmadvand, L.~Bian and S.~Shakeri, \emph{{Heavy QCD axion model in light of pulsar timing arrays}}, \href{http://dx.doi.org/10.1103/PhysRevD.108.115020}{\emph{Phys. Rev. D} {\bf 108} (2023) 115020} [\href{https://arxiv.org/abs/2307.12385}{{\tt arXiv:2307.12385}}].

\bibitem{Bedi:2024wqg}
R.~Bedi, T.~Gherghetta, C.~Grojean, G.~Guedes, J.~Kley and P.~N.~H. Vuong, \emph{{Small instanton-induced flavor invariants and the axion potential}}, \href{http://dx.doi.org/10.1007/JHEP06(2024)156}{\emph{JHEP} {\bf 06} (2024) 156} [\href{https://arxiv.org/abs/2402.09361}{{\tt arXiv:2402.09361}}].

\bibitem{Chakraborty:2024tyx}
S.~Chakraborty, A.~Gupta and M.~Vanvlasselaer, \emph{{Photoproduction of heavy QCD axions in supernovae}}, \href{http://dx.doi.org/10.1103/PhysRevD.110.063032}{\emph{Phys. Rev. D} {\bf 110} (2024) 063032} [\href{https://arxiv.org/abs/2403.12169}{{\tt arXiv:2403.12169}}].

\bibitem{EuropeanStrategyforParticlePhysicsPreparatoryGroup:2019qin}
R.~K. Ellis et~al., \emph{{Physics Briefing Book}: {Input for the European Strategy for Particle Physics Update 2020}},  \href{https://arxiv.org/abs/1910.11775}{{\tt arXiv:1910.11775}}.

\bibitem{Davighi:2022qgb}
J.~Davighi, A.~Greljo and A.~E. Thomsen, \emph{{Leptoquarks with exactly stable protons}}, \href{http://dx.doi.org/10.1016/j.physletb.2022.137310}{\emph{Phys. Lett. B} {\bf 833} (2022) 137310} [\href{https://arxiv.org/abs/2202.05275}{{\tt arXiv:2202.05275}}].

\bibitem{Davighi:2022fer}
J.~Davighi and J.~Tooby-Smith, \emph{{Electroweak flavour unification}}, \href{http://dx.doi.org/10.1007/JHEP09(2022)193}{\emph{JHEP} {\bf 09} (2022) 193} [\href{https://arxiv.org/abs/2201.07245}{{\tt arXiv:2201.07245}}].

\bibitem{Davighi:2022vpl}
J.~Davighi, \emph{{Gauge flavour unification: From the flavour puzzle to stable protons}}, \href{http://dx.doi.org/10.1393/ncc/i2023-23023-0}{\emph{Nuovo Cim. C} {\bf 46} (2022) 23} [\href{https://arxiv.org/abs/2206.04482}{{\tt arXiv:2206.04482}}].

\bibitem{Davighi:2023xqn}
J.~Davighi, A.~Gosnay, D.~J. Miller and S.~Renner, \emph{{Phenomenology of a Deconstructed Electroweak Force}}, \href{http://dx.doi.org/10.1007/JHEP05(2024)085}{\emph{JHEP} {\bf 05} (2024) 085} [\href{https://arxiv.org/abs/2312.13346}{{\tt arXiv:2312.13346}}].

\end{thebibliography}\endgroup

\end{document}